\documentclass[prb,twocolumn,superscriptaddress,showpacs]{revtex4}
\usepackage{amsmath}
\usepackage{amsfonts}
\usepackage{graphicx}
\usepackage{pst-all}

\begin{document}

\title{Measurement-Only Topological Quantum Computation via Tunable Interactions}
\author{Parsa Bonderson}
\affiliation{Station Q, Microsoft Research, Santa Barbara, California 93106-6105, USA}
\date{\today}

\begin{abstract}
I examine, in general, how tunable interactions may be used to perform anyonic teleportation and generate braiding transformations for non-Abelian anyons. I explain how these methods are encompassed by the ``measurement-only'' approach to topological quantum computation. The physically most relevant example of Ising anyons or Majorana zero modes is considered in detail, particularly in the context of Majorana nanowires.
\end{abstract}

\maketitle









%

\section{Introduction}

Non-Abelian anyons are quasiparticle excitations of topological phases that exhibit exotic exchange statistics governed by higher-dimensional representations of the braid group~\cite{Leinaas77,Goldin85,Fredenhagen89,Froehlich90}. Such quasiparticles collectively possess a multi-dimensional, non-local (topological) state space that is essentially immune to local perturbations. This property makes non-Abelian topological phases appealing platforms for quantum information processing, as they allow for topologically protected quantum computation (TQC)~\cite{Kitaev03,Nayak08}. In the TQC approach, computational gates may be generated through topological operations, such as braiding exchanges of quasiparticles, in which case they are also topologically protected.
The physical implementation of such protected gates poses one of the most significant challenges for realization of TQC.

The initial conception of TQC envisioned physically translocating non-Abelian quasiparticles to perform braiding operations as the primary means of generating gates. Proposals for moving quasiparticles include simply dragging them around (e.g., with a STM tip, if they are electrically charged) and a ``bucket brigade'' series of induced hoppings from one site to the next, originating at one location and terminating at another~\cite{Freedman06a}.
A subsequent proposal, known as ``measurement-only TQC'' (MOTQC)~\cite{Bonderson08a,Bonderson08b}, introduced methods of effectively generating braiding transformations on the state space, without physically moving the anyons associated with the state space. These transformations are implemented by executing a series of measurements and anyonic state teleportations.

Recently, there have been a number of proposals to utilize coupling interactions of some sort, e.g., topological charge tunneling or Coulomb interactions, which are used to replace (or supplement) the physical translocation of non-Abelian anyons and implement braiding operations on their non-local state space~\cite{Sau11a,vanHeck12a,Clarke12,Lindner12,Barkeshli12,Burrello12}. I consider these interaction-based proposals in generality and explain how this class of methods is encompassed by the MOTQC approach. I also examine the physically most relevant example of Ising anyons or Majorana zero modes in detail, particularly in the context of Majorana wires.

\section{Formalism}

In this paper, I will use the diagrammatic representation of anyonic states and operators, as described by a general anyon model. This encodes the purely topological properties of quasiparticles, independent of any particular physical realization. For additional details and conventions used in this paper, I refer the reader to Refs.~\onlinecite{Bonderson07b,Bonderson07c,Bonderson08b}.

An anyon model is defined by a set $\mathcal{C}$ of conserved quantum numbers called topological charge, fusion rules specifying what can result from combining or splitting topological charges, and braiding rules specifying what happens when the positions of objects carrying topological charge are exchanged. Each quasiparticle carries a definite localized value of topological charge. There is a unique ``vacuum'' charge, denoted $I$, for which fusion and braiding is trivial, and each charge $a$ has a unique conjugate $\bar{a}$ which can fuse with $a$ to give $I$. The topological charges obey the anyon model's (commutative and associative) fusion algebra
\begin{equation}
a \times b = \sum_{c} N_{ab}^{c} c
,
\end{equation}
and where $N_{ab}^{c}$ are non-negative integers specifying the number of ways that topological charges $a$ and $b$ can combine to produce charge $c$.
These rules prescribe fusion/splitting Hilbert spaces $\mathcal{V}_{ab}^{c}$ and $\mathcal{V}^{ab}_{c}$ with $\dim(\mathcal{V}_{ab}^{c}) = \dim(\mathcal{V}^{ab}_{c}) = N_{ab}^{c}$, which generate the non-local state space through repeated fusion/splitting. A charge $a$ is non-Abelian if it does not have a unique fusion channel for every type of charge it is fused with, or, alternatively, if it has $\sum_{c} N^{c}_{aa} > 1$. It is clear that the dimension of the topological state space increases as one includes more non-Abelian anyons.

Diagrammatically, the orthonormal bra/ket vectors in the fusion/splitting spaces are represented by trivalent vertices:
\begin{eqnarray}
\left( d_{c} / d_{a}d_{b} \right) ^{1/4}
\pspicture[shift=-0.6](-0.1,-0.2)(1.5,-1.2)
  \small
  \psset{linewidth=0.9pt,linecolor=black,arrowscale=1.5,arrowinset=0.15}
  \psline{-<}(0.7,0)(0.7,-0.35)
  \psline(0.7,0)(0.7,-0.55)
  \psline(0.7,-0.55) (0.25,-1)
  \psline{-<}(0.7,-0.55)(0.35,-0.9)
  \psline(0.7,-0.55) (1.15,-1)	
  \psline{-<}(0.7,-0.55)(1.05,-0.9)
  \rput[tl]{0}(0.4,0){$c$}
  \rput[br]{0}(1.4,-0.95){$b$}
  \rput[bl]{0}(0,-0.95){$a$}
 \scriptsize
  \rput[bl]{0}(0.85,-0.5){$\mu$}
  \endpspicture
&=&\left\langle a,b;c,\mu \right| \in
\mathcal{V}_{ab}^{c} ,
\label{eq:bra}
\\
\left( d_{c} / d_{a}d_{b}\right) ^{1/4}
\pspicture[shift=-0.65](-0.1,-0.2)(1.5,1.2)
  \small
  \psset{linewidth=0.9pt,linecolor=black,arrowscale=1.5,arrowinset=0.15}
  \psline{->}(0.7,0)(0.7,0.45)
  \psline(0.7,0)(0.7,0.55)
  \psline(0.7,0.55) (0.25,1)
  \psline{->}(0.7,0.55)(0.3,0.95)
  \psline(0.7,0.55) (1.15,1)	
  \psline{->}(0.7,0.55)(1.1,0.95)
  \rput[bl]{0}(0.4,0){$c$}
  \rput[br]{0}(1.4,0.8){$b$}
  \rput[bl]{0}(0,0.8){$a$}
 \scriptsize
  \rput[bl]{0}(0.85,0.35){$\mu$}
  \endpspicture
&=&\left| a,b;c,\mu \right\rangle \in
\mathcal{V}_{c}^{ab},
\label{eq:ket}
\end{eqnarray}
where $\mu = 1, \ldots, N_{ab}^{c}$. The normalization factors involving $d_{a}$, the quantum dimension of the charge $a$, are included so that diagrams are in the isotopy-invariant convention. States and operators involving multiple anyons are constructed by appropriately stacking together diagrams, making sure to conserve charge when connecting endpoints of lines. In this paper, I consider only anyon models with no fusion multiplicities, i.e., $N_{ab}^{c}=0$ or $1$, (which includes all the cases of physical interest,) and so will leave the basis labels $\mu$ implicit in the rest of the paper, but the discussion may be generalized.

Associativity of fusion in the state space is encoded by the unitary (change of fusion basis) isomorphisms
$F^{abc}_{d} : \bigoplus_{e} \mathcal{V}^{ab}_{e} \otimes \mathcal{V}^{ec}_{d} \rightarrow \bigoplus_{e} \mathcal{V}^{af}_{d} \otimes \mathcal{V}^{bc}_{f}$. These $F$-symbols are similar to the $6j$-symbols of angular momentum representations. Diagrammatically, these are written as
\begin{equation}
\label{eq:F}
\psscalebox{1}{
\pspicture[shift=-1.0](0,-0.4)(1.8,1.8)
  \small
  \psset{linewidth=0.9pt,linecolor=black,arrowscale=1.5,arrowinset=0.15}
  \psline(1,0.5)(1,0)
  \psline(0.2,1.5)(1,0.5)
  \psline(1.8,1.5) (1,0.5)
  \psline(0.6,1) (1,1.5)
   \psline{->}(0.6,1)(0.3,1.375)
   \psline{->}(0.6,1)(0.9,1.375)
   \psline{->}(1,0.5)(1.7,1.375)
   \psline{->}(1,0.5)(0.7,0.875)
   \psline{->}(1,0)(1,0.375)
   \rput[bl]{0}(0.05,1.6){$a$}
   \rput[bl]{0}(0.95,1.6){$b$}
   \rput[bl]{0}(1.75,1.6){$c$}
   \rput[bl]{0}(0.5,0.5){$e$}
   \rput[bl]{0}(.95,-0.3){$d$}
  \endpspicture
}
= \sum_{f} \left[F_{d}^{abc}\right]_{ef}
\psscalebox{1}{
\pspicture[shift=-1.0](0.2,-0.4)(1.8,1.8)
  \small
  \psset{linewidth=0.9pt,linecolor=black,arrowscale=1.5,arrowinset=0.15}
  \psline(1,0.5)(1,0)
  \psline(0.2,1.5)(1,0.5)
  \psline(1.8,1.5) (1,0.5)
  \psline(1.4,1) (1,1.5)
   \psline{->}(0.6,1)(0.3,1.375)
   \psline{->}(1.4,1)(1.1,1.375)
   \psline{->}(1,0.5)(1.7,1.375)
   \psline{->}(1,0.5)(1.3,0.875)
   \psline{->}(1,0)(1,0.375)
   \rput[bl]{0}(0.05,1.6){$a$}
   \rput[bl]{0}(0.95,1.6){$b$}
   \rput[bl]{0}(1.75,1.6){$c$}
   \rput[bl]{0}(1.25,0.45){$f$}
   \rput[bl]{0}(.95,-0.3){$d$}
  \endpspicture
}
.
\end{equation}

The counterclockwise braiding exchange operator of topological charges $a$ and $b$ is represented diagrammatically by
\begin{equation}
\label{eq:R}
R_{ab}=
\pspicture[shift=-0.6](-0.1,-0.2)(1.3,1.05)
\small
  \psset{linewidth=0.9pt,linecolor=black,arrowscale=1.5,arrowinset=0.15}
  \psline(0.96,0.05)(0.2,1)
  \psline{->}(0.96,0.05)(0.28,0.9)
  \psline(0.24,0.05)(1,1)
  \psline[border=2pt]{->}(0.24,0.05)(0.92,0.9)
  \rput[bl]{0}(-0.02,0){$a$}
  \rput[br]{0}(1.2,0){$b$}
  \endpspicture
=\sum\limits_{c}\sqrt{\frac{d_{c}}{d_{a}d_{b}}}
\, R_{c}^{ab}
\pspicture[shift=-1.1](0.1,-0.85)(1.3,1.3)
 \small
  \psset{linewidth=0.9pt,linecolor=black,arrowscale=1.5,arrowinset=0.15}
  \psline{->}(0.7,0)(0.7,0.45)
  \psline(0.7,0)(0.7,0.55)
  \psline(0.7,0.55) (0.2,1.05)
  \psline{->}(0.7,0.55)(0.3,0.95)
  \psline(0.7,0.55) (1.2,1.05)
  \psline{->}(0.7,0.55)(1.1,0.95)
  \rput[bl]{0}(0.88,0.2){$c$}
  \rput[bl]{0}(1.1,1.15){$a$}
  \rput[bl]{0}(0.1,1.15){$b$}
  \psline(0.7,0) (0.2,-0.5)
  \psline{-<}(0.7,0)(0.35,-0.35)
  \psline(0.7,0) (1.2,-0.5)
  \psline{-<}(0.7,0)(1.05,-0.35)
  \rput[bl]{0}(1.1,-0.8){$b$}
  \rput[bl]{0}(0.1,-0.8){$a$}
  \endpspicture
\end{equation}
where the symbols $R_{c}^{ab}$ are the phases acquired by exchanging anyons of charge $a$ and $b$, which fuse together into fusion channel $c$. Similarly, the clockwise braid is
\begin{equation}
R_{ab}^{\dag}= R_{ab}^{-1}=
\pspicture[shift=-0.6](-0.1,-0.2)(1.3,1.05)
\small
  \psset{linewidth=0.9pt,linecolor=black,arrowscale=1.5,arrowinset=0.15}
  \psline{->}(0.24,0.05)(0.92,0.9)
  \psline(0.24,0.05)(1,1)
  \psline(0.96,0.05)(0.2,1)
  \psline[border=2pt]{->}(0.96,0.05)(0.28,0.9)
  \rput[bl]{0}(-0.01,0){$b$}
  \rput[bl]{0}(1.06,0){$a$}
  \endpspicture
.
\end{equation}

\begin{figure}[t!]
\begin{center}
  \includegraphics[scale=0.11]{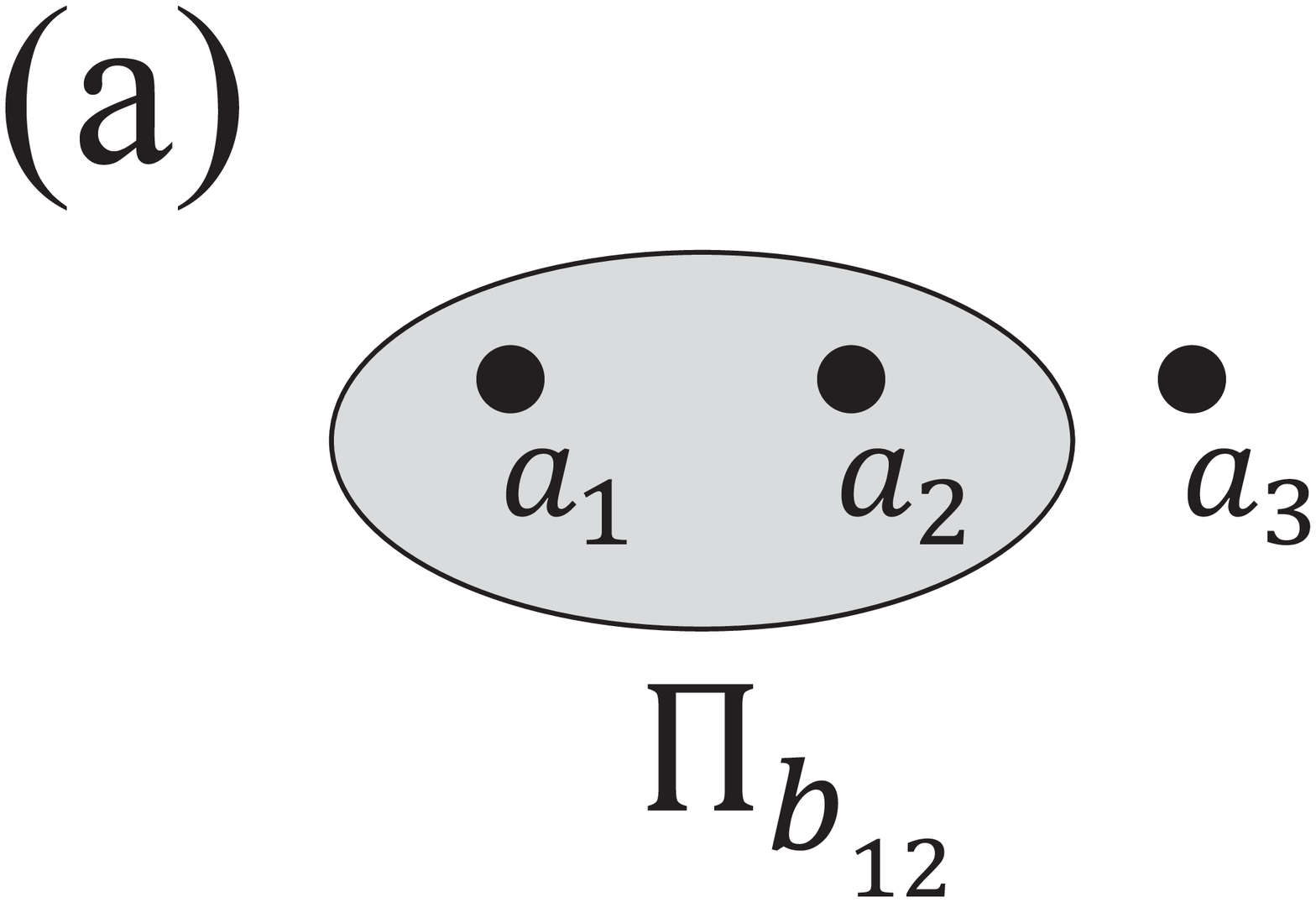}
  \includegraphics[scale=0.11]{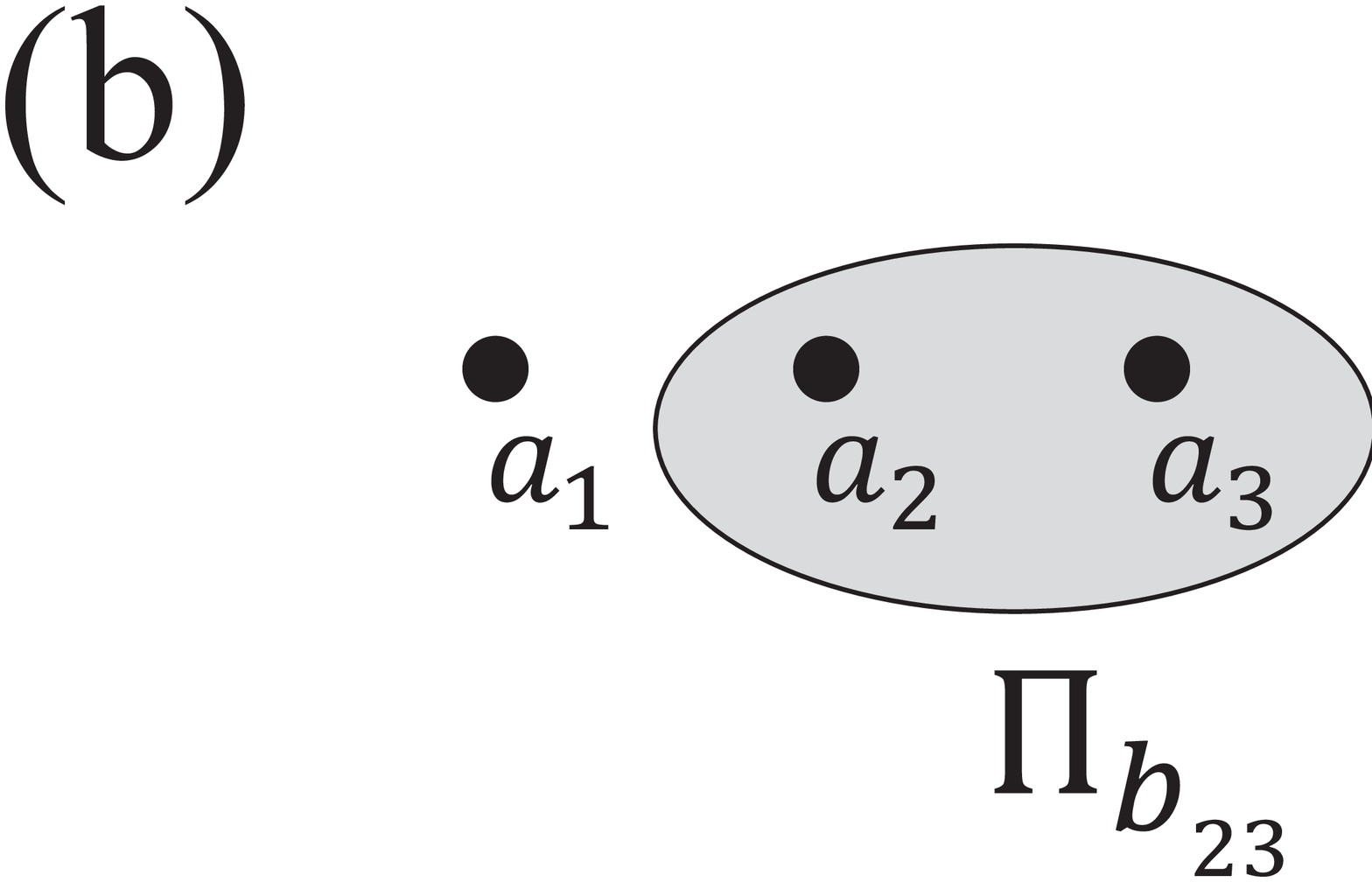}
  \includegraphics[scale=0.11]{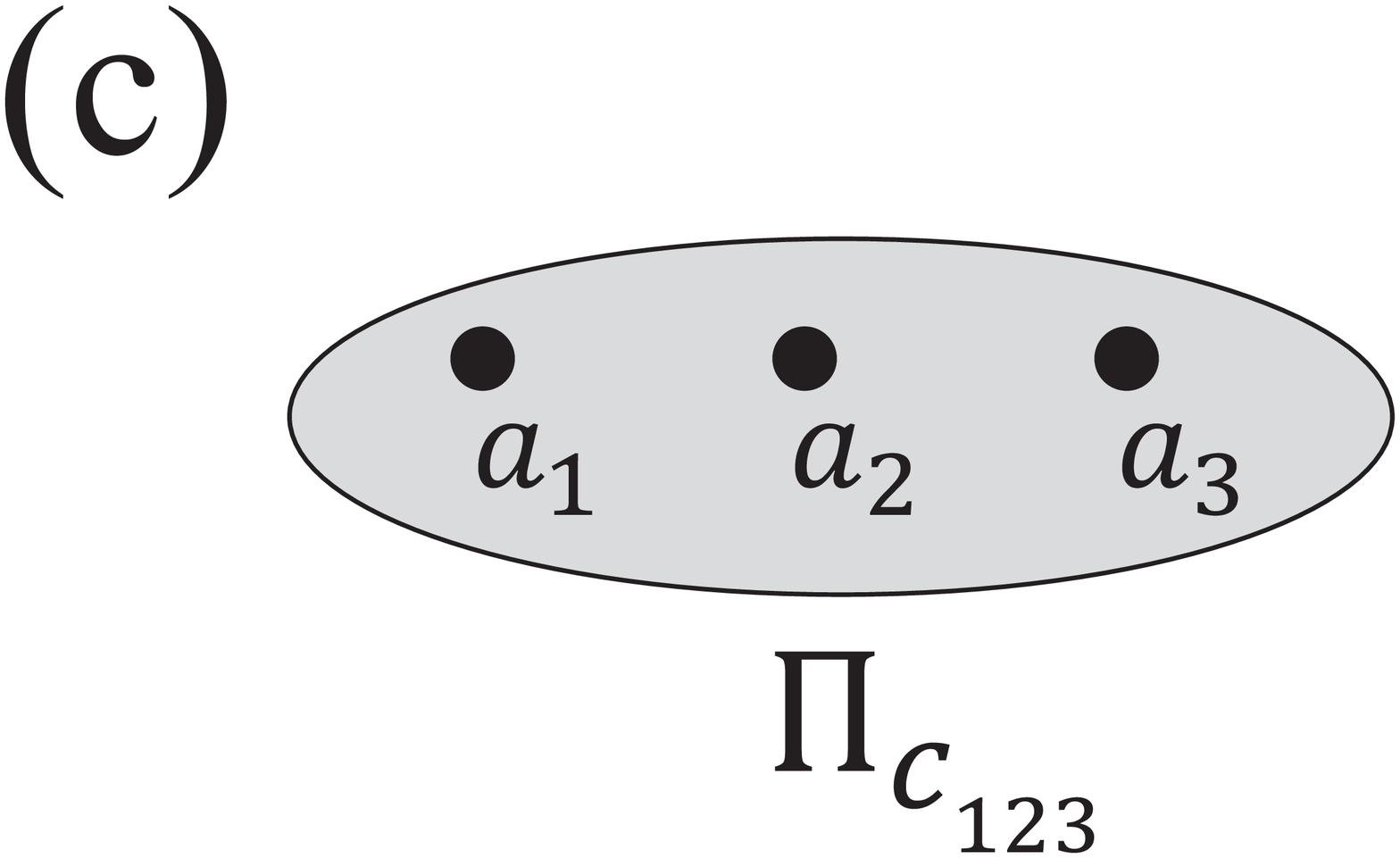}
  \caption{Topological charge projectors (indicated by the shaded ovals) occurring for (a) anyons $1$ and $2$, (b) anyons $2$ and $3$, and (c) anyons $1$, $2$, and $3$.}
  \label{fig:projectors}
\end{center}
\end{figure}

The projection of two anyons with topological charges $a_1$ and $a_2$, respectively, onto collective topological charge $b$ is given by
\begin{equation}
\Pi^{(12)}_{b} = \sqrt{\frac{d_{b}} {d_{a_1} d_{a_2} }}
\pspicture[shift=-1.1](-0.1,-0.85)(1.6,1.3)
 \small
  \psset{linewidth=0.9pt,linecolor=black,arrowscale=1.5,arrowinset=0.15}
  \psline{->}(0.7,0)(0.7,0.45)
  \psline(0.7,0)(0.7,0.55)
  \psline(0.7,0.55) (0.2,1.05)
  \psline{->}(0.7,0.55)(0.3,0.95)
  \psline(0.7,0.55) (1.2,1.05)
  \psline{->}(0.7,0.55)(1.1,0.95)
  \rput[bl]{0}(0.88,0.2){$b$}
  \rput[bl]{0}(1.1,1.15){$a_{2}$}
  \rput[bl]{0}(0.1,1.15){$a_{1}$}
  \psline(0.7,0) (0.2,-0.5)
  \psline{-<}(0.7,0)(0.35,-0.35)
  \psline(0.7,0) (1.2,-0.5)
  \psline{-<}(0.7,0)(1.05,-0.35)
  \rput[bl]{0}(1.1,-0.8){$a_{2}$}
  \rput[bl]{0}(0.1,-0.8){$a_{1}$}
  \endpspicture
.
\end{equation}
The projection of three anyons with topological charges $a_1$, $a_2$, and $a_3$, respectively, onto collective topological charge $c$ is given by
\begin{equation}
\Pi^{(123)}_{c} = \sum_{b} \sqrt{\frac{d_{c}} {d_{a_1} d_{a_2} d_{a_3} }}
\pspicture[shift=-1.7](-1.2,-1.5)(1.2,1.8)
  \small
  \psset{linewidth=0.9pt,linecolor=black,arrowscale=1.5,arrowinset=0.15}
  \psline(0.0,0.5)(0.8,1.5)
  \psline(0.0,0.5)(-0.8,1.5)
  \psline(-0.4,1)(0.0,1.5)
    \psline{->}(-0.4,1)(-0.1,1.375)
    \psline{->}(0.4,1.0)(0.7,1.375)
    \psline{->}(0,0.5)(-0.3,0.875)
    \psline{->}(0,0.5)(-0.7,1.375)
  \psset{linewidth=0.9pt,linecolor=black,arrowscale=1.5,arrowinset=0.15}
  \psline(0.0,0.0)(0.8,-1)
  \psline(0.0,0.0)(-0.8,-1)
  \psline(-0.4,-0.5)(0.0,-1)
    \psline{-<}(-0.4,-0.5)(-0.1,-0.875)
    \psline{-<}(0,0.0)(0.7,-0.875)
    \psline{-<}(0,0.0)(-0.3,-0.375)
    \psline{-<}(0,0.0)(-0.7,-0.875)
  \psline(0,0.0)(0,0.5)
  \psline{->}(0,0.0)(0,0.4)
  \rput[bl]{0}(0.15,0.15){$c$}
  \rput[bl]{0}(-0.55,0.5){$b$}
  \rput[bl]{0}(-0.95,1.6){$a_{1}$}
  \rput[bl]{0}(-0.1,1.6){$a_{2}$}
  \rput[bl]{0}(0.7,1.6){$a_{3}$}
  \rput[bl]{0}(-0.55,-0.3){$b$}
  \rput[bl]{0}(-0.95,-1.4){$a_{1}$}
  \rput[bl]{0}(-0.1,-1.4){$a_{2}$}
  \rput[bl]{0}(0.7,-1.4){$a_{3}$}
\endpspicture
.
\end{equation}
The planar representation of two and three anyon projectors are shown in Fig.~\ref{fig:projectors}. When an operator acts on only a subset of all the anyons, it implicitly means that it acts trivially on the other anyons, e.g. $\Pi^{(12)}_{b}$ really means $\Pi^{(12)}_{b} \otimes \openone^{(3 \ldots n)}$ when there are $n$ anyons.

\begin{figure}[t!]
\begin{center}
  \includegraphics[scale=0.11]{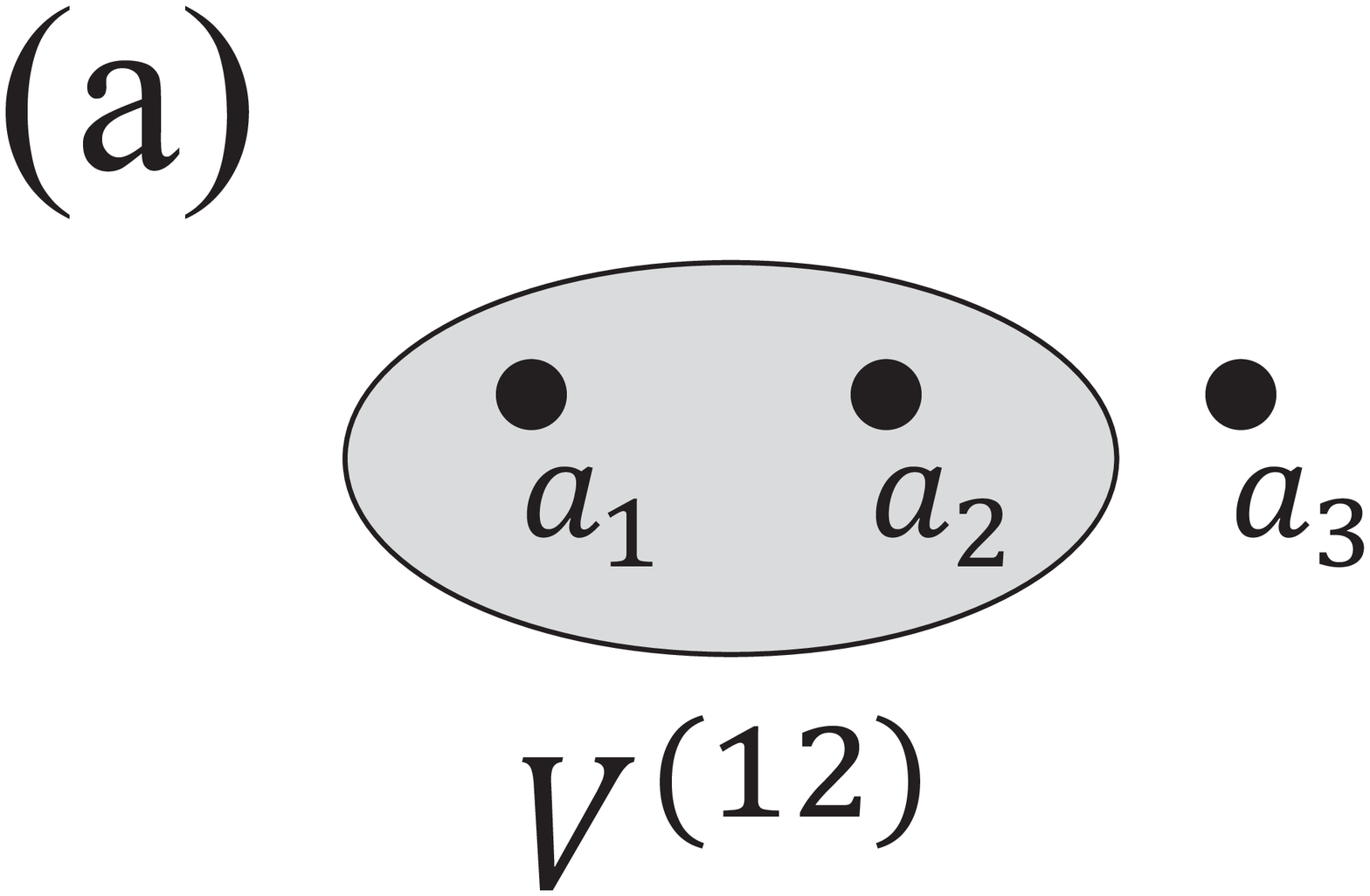}
  \includegraphics[scale=0.11]{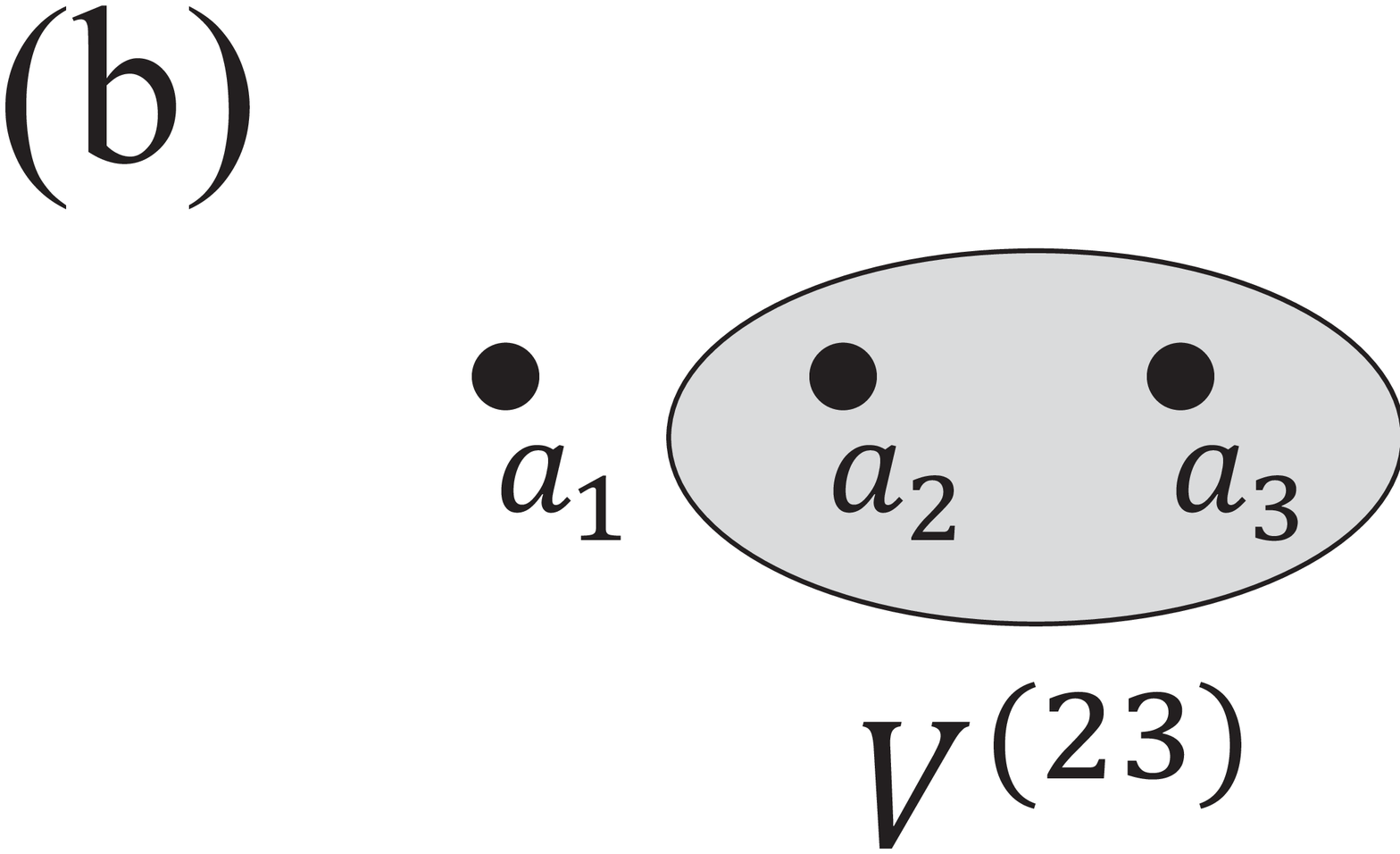}
  \includegraphics[scale=0.11]{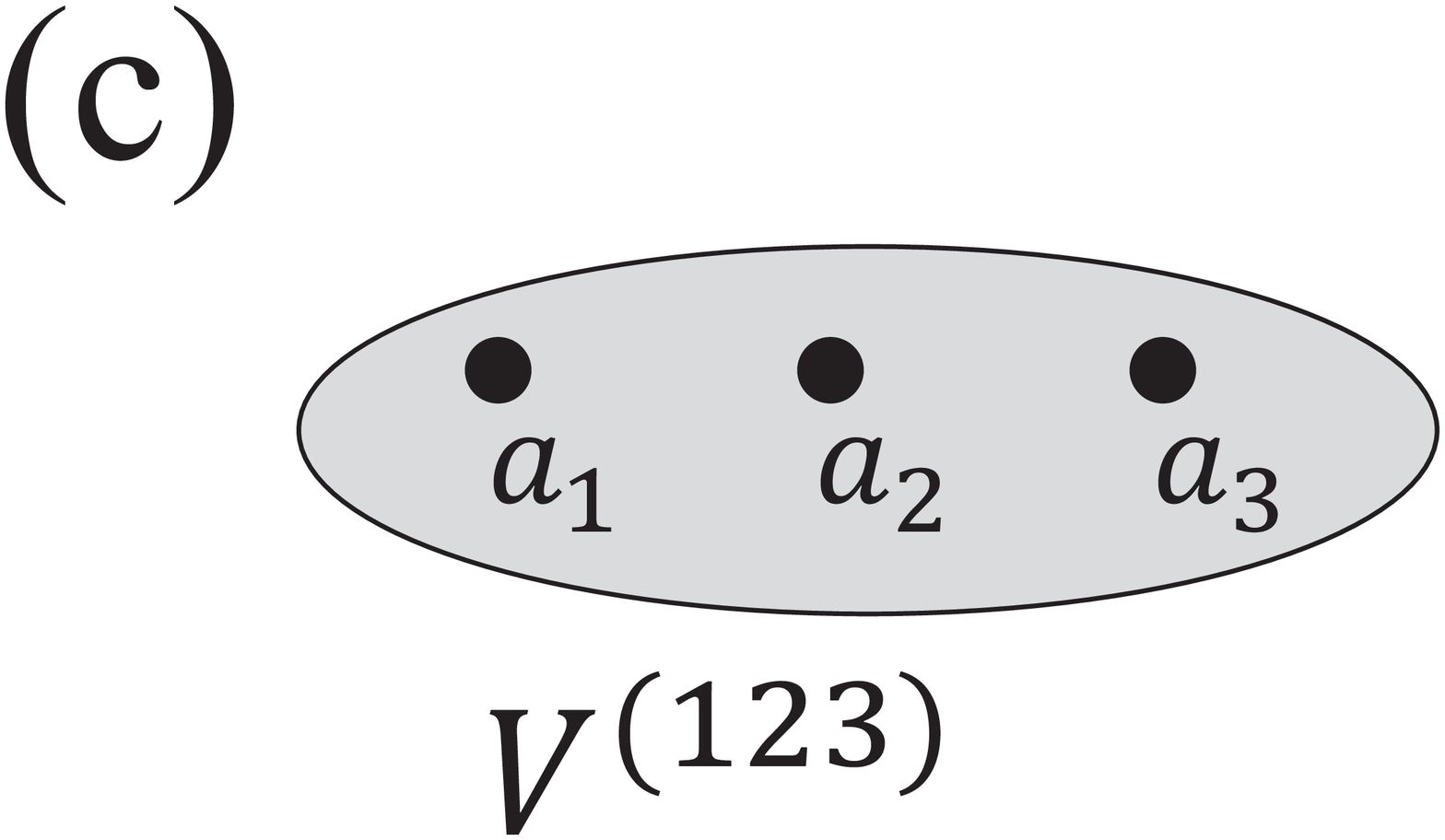}
  \caption{Interactions (indicated by the shaded ovals) occurring (a) pairwise between anyons $1$ and $2$, (b) pairwise between anyons $2$ and $3$, and (c) between three anyons $1$, $2$, and $3$.}
  \label{fig:interactions}
\end{center}
\end{figure}

A general pairwise interaction between two anyons can be expressed in terms of tunneling of topological charge between the two anyons as
\begin{equation}
\label{eq:V1}
V^{(12)} = \sum_{e} \left( \Gamma_{e} \frac{1}{\sqrt{d_{e}}}
\pspicture[shift=-1.1](0.1,-1.2)(1.5,1.15)
  \small
  \psset{linewidth=0.9pt,linecolor=black,arrowscale=1.5,arrowinset=0.15}
  \psline(0.35,-0.775)(0.35,0.775)
  \psline{>-}(0.35,0.3)(0.35,0.6)
  \psline{<-}(0.35,-0.3)(0.35,-0.6)
  \psline(1.35,-0.775)(1.35,0.775)
  \psline{>-}(1.35,0.3)(1.35,0.6)
  \psline{<-}(1.35,-0.3)(1.35,-0.6)
  \psline(1.35,0.1)(0.35,-0.1)
   \psline{->}(0.35,-0.1)(1,0.03)
  \rput[bl]{0}(0.25,0.875){$a_1$}
  \rput[bl]{0}(1.25,0.875){${a_2}$}
  \rput[bl]{0}(0.25,-1.075){$a_1$}
  \rput[bl]{0}(1.25,-1.075){${a_2}$}
  \rput[br]{0}(0.95,0.13){$e$}
\endpspicture
+\Gamma_{e}^{\ast} \frac{1}{\sqrt{d_{e}}}
\pspicture[shift=-1.1](0.1,-1.2)(1.5,1.15)
  \small
  \psset{linewidth=0.9pt,linecolor=black,arrowscale=1.5,arrowinset=0.15}
  \psline(0.35,-0.775)(0.35,0.775)
  \psline{>-}(0.35,0.3)(0.35,0.6)
  \psline{<-}(0.35,-0.3)(0.35,-0.6)
  \psline(1.35,-0.775)(1.35,0.775)
  \psline{>-}(1.35,0.3)(1.35,0.6)
  \psline{<-}(1.35,-0.3)(1.35,-0.6)
  \psline(0.35,0.1)(1.35,-0.1)
   \psline{->}(1.35,-0.1)(0.7,0.03)
  \rput[bl]{0}(0.25,0.875){$a_1$}
  \rput[bl]{0}(1.25,0.875){${a_2}$}
  \rput[bl]{0}(0.25,-1.075){$a_1$}
  \rput[bl]{0}(1.25,-1.075){${a_2}$}
  \rput[br]{0}(0.95,0.13){$e$}
\endpspicture
\right)
,
\end{equation}
where $\Gamma_e$ is the tunneling amplitude of charge $e$, or in terms of projectors as
\begin{equation}
V^{(12)} = \sum_{b} E_{b} \, \Pi_{b}^{(12)}
\end{equation}
where $E_{b}$ is the energy associated with the two anyons having fusion channel $b$. These are related by
\begin{equation}
\label{eq:E_to_Gamma}
E_{b} = \sum_{e} \left(  \Gamma_{e} \left[ F^{a_1 e a_2}_{b} \right]_{ a_1 a_2 } + \Gamma_{e}^{\ast} \left[ F^{a_1 e a_2}_{b} \right]_{ a_1 a_2 }^{\ast} \right),
\end{equation}
and tunneling will generically fully split the fusion channel degeneracy of a pair of non-Abelian anyons~\cite{Bonderson09}.

$n$-anyon interactions can similarly be defined, but they will generally include tunneling terms or projectors that involve up to $n$ anyons. These will not be very important in this paper, so I will not go into detail. The planar representation of two and three anyon interactions are shown in Fig.~\ref{fig:interactions}. The resemblance to the representation of projectors is intended to emphasize their close relation.

No interaction can truly be turned off to exactly zero (except with fine-tuning), since a physical system has finite size and separations between quasiparticles. However, it is generally possible to exponentially suppress interactions in topological phases, for example, by separating quasiparticles by distances much greater than the correlation length. I will make no further distinction between such exponentially suppressed quantities and zero.

\section{Generating Forced Measurements Using Tunable Interactions}
\label{sec:Interaction_Forced_Measurement}

In this section, I demonstrate how adiabatic manipulation of interactions between anyons may be used to implement certain topological charge projection operators, such as those used for anyonic teleportation and MOTQC. This can be done by restricting one's attention to three non-Abelian anyons that carry charges $a_1$, $a_2$, and $a_3$, respectively, and have definite collective topological charge $c$, which is non-Abelian. The internal fusion state space of these three anyons is $\mathcal{V}^{a_1 a_2 a_3}_{c} \cong \bigoplus_{e} \mathcal{V}^{a_1 a_2}_{e} \otimes \mathcal{V}^{e a_3}_{c} \cong \bigoplus_{f} \mathcal{V}^{a_1 f}_{c} \otimes \mathcal{V}^{a_2 a_3}_{f}$.

Consider a time-dependent Hamiltonian $H \left( t \right)$ with the following properties:

1. $H \left( 0 \right) =  V^{(23)}$ is an interaction involving only anyons $2$ and $3$, for which the ground states have definite topological charge value $b_{23}$ for the fusion channel of anyons $2$ and $3$ (i.e. they are eigenstates of the projector $\Pi^{(23)}_{b_{23}}$ that survive projection) and there is an energy gap to states with other topological charge values of this fusion channel.

2. $H \left( \tau \right) =  V^{(12)}$ is an interaction involving only anyons $1$ and $2$, for which the ground states have definite topological charge value $b_{12}$ for the fusion channel of anyons $1$ and $2$ (i.e. they are eigenstates of the projector $\Pi^{(12)}_{b_{12}}$ that survive projection) and there is an energy gap to states with other topological charge values of this fusion channel.

3. $H \left( t \right)$ is an interaction involving only anyons $1$, $2$, and $3$, which adiabatically connects $H \left( 0 \right)$ and $H \left( \tau \right)$, without closing the gap between the ground states and the higher energy states, during $0 \leq t \leq \tau$.

In other words, the ground state subspace of $H(t)$ corresponds to a one-dimensional subspace of $\mathcal{V}^{a_1 a_2 a_3}_{c}$ for all $0 \leq t \leq \tau$, where this subspace is $\mathcal{V}^{a_1 b_{23}}_{c} \otimes \mathcal{V}^{a_2 a_3}_{b_{23}}$ at $t=0$ and $\mathcal{V}^{a_1 a_2}_{b_{12}} \otimes \mathcal{V}^{b_{12} a_3}_{c}$ at $t=\tau$. I emphasize that, even though the internal fusion channel degeneracy of anyons $1$, $2$, and $3$ is broken, $H \left( t \right)$ may still exhibit ground state degeneracy due to the any other anyons in the system, since it acts trivially upon them. Also, since $H \left( t \right)$ is an interaction involving only anyons $1$, $2$, and $3$, it cannot change their collective topological charge $c$.

It is now easy to apply the adiabatic theorem to determine the result of (unitary) time evolution on the ground state subspace. The adiabatic theorem states that if the system is in an energy eigenstate and it goes through an adiabatic process which does not close the gap between the corresponding instantaneous energy eigenvalue and the rest of the Hamiltonian's spectrum, then the system will remain in the subspace corresponding to this instantaneous energy eigenvalue. Since the Hamiltonian only acts nontrivially on anyons $1$, $2$, and $3$, and a ground state will stay in the instantaneous ground state subspace, the resulting ground state evolution operator $U_{0}(t)$ [i.e., the restriction of the time evolution operator $U(t)$ to the ground state subspace] at time $t=\tau$ must be
\begin{eqnarray}
U_{0} (\tau) &=& e^{i \theta} \left( \left[ F^{a_{1} a_{2} a_{3} }_{c} \right]^{\ast}_{b_{12} b_{23}} \right)^{-1} \Pi^{(12)}_{b_{12}} \,\, \Pi^{(123)}_{c} \,\, \Pi^{(23)}_{b_{23}} \notag \\
&=& e^{i \theta} \left| a_1 , a_2 ; b_{12} \right\rangle \left| b_{12} , a_{3} ; c \right\rangle  \left\langle a_{1}, b_{23} ; c \right| \left\langle a_2 , a_3 ; b_{23} \right|  \notag \\
&=& e^{i \theta} \sqrt{\frac{d_c}{d_{a_1} d_{a_2} d_{a_3} }}
\pspicture[shift=-1.7](-1.2,-1.6)(1.2,2.0)
  \small
  \psset{linewidth=0.9pt,linecolor=black,arrowscale=1.5,arrowinset=0.15}
  \psline(0.0,0.5)(0.8,1.5)
  \psline(0.0,0.5)(-0.8,1.5)
  \psline(-0.4,1)(0.0,1.5)
    \psline{->}(-0.4,1)(-0.1,1.375)
    \psline{->}(0.4,1.0)(0.7,1.375)
    \psline{->}(0,0.5)(-0.3,0.875)
    \psline{->}(0,0.5)(-0.7,1.375)
  \psset{linewidth=0.9pt,linecolor=black,arrowscale=1.5,arrowinset=0.15}
  \psline(0.0,0.0)(0.8,-1)
  \psline(0.0,0.0)(-0.8,-1)
  \psline(0.4,-0.5)(0.0,-1)
    \psline{-<}(0.4,-0.5)(0.1,-0.875)
    \psline{-<}(0,0.0)(0.7,-0.875)
    \psline{-<}(0,0.0)(0.3,-0.375)
    \psline{-<}(0,0.0)(-0.7,-0.875)
  \psline(0,0.0)(0,0.5)
  \psline{->}(0,0.0)(0,0.4)
  \rput[bl]{0}(0.15,0.05){$c$}
  \rput[bl]{0}(-0.7,0.5){$b_{12}$}
  \rput[bl]{0}(-0.95,1.6){$a_{1}$}
  \rput[bl]{0}(-0.1,1.6){$a_{2}$}
  \rput[bl]{0}(0.7,1.6){$a_{3}$}
  \rput[bl]{0}(0.4,-0.35){$b_{23}$}
  \rput[bl]{0}(-0.95,-1.4){$a_{1}$}
  \rput[bl]{0}(-0.1,-1.4){$a_{2}$}
  \rput[bl]{0}(0.7,-1.4){$a_{3}$}
 \endpspicture
\end{eqnarray}
where the factor $\left[ F^{a_{1} a_{2} a_{3} }_{c} \right]^{\ast}_{b_{12} b_{23}}$ is necessary to ensure that the operator is unitary. The (unimportant) overall phase $e^{i \theta}$ is the product of the dynamical phase and the Berry's phase. I note that $\Pi^{(123)}_{c}$ commutes with $\Pi^{(12)}_{b_{12}}$ and $\Pi^{(23)}_{b_{23}}$.

\begin{figure}[t!]
\begin{center}
  \includegraphics[scale=0.3]{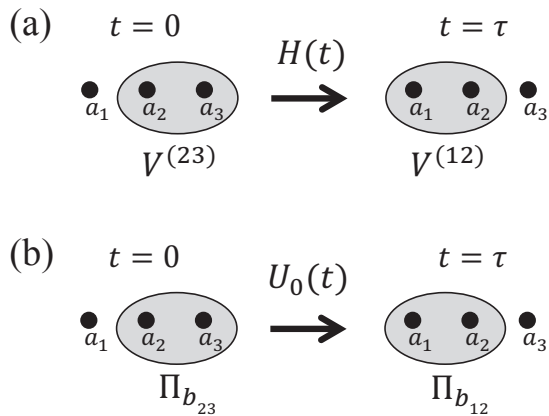}
  \caption{(a) A Hamiltonian $H(t)$ which adiabatically connects the interaction $V^{(23)}$ with energetically preferred fusion channel $b_{23}$ to the interaction $V^{(12)}$ with energetically preferred fusion channel $b_{12}$, without closing the gap. (b) This Hamiltonian implements time evolution $U_{0}(t)$ on the ground state space which takes eigenstates of the projector $\Pi^{(23)}_{b_{23}}$ that survive projection to eigenstates of the projector $\Pi^{(12)}_{b_{12}}$ that survive projection. This has the same effect as applying a projector $\Pi^{(12)}_{b_{12}}$ to an initial ground state, up to normalizing factors and an unimportant overall phase.}
  \label{fig:proj_via_int}
\end{center}
\end{figure}

Thus, it is clear that applying the operator $U_{0} (\tau)$ to states in the $t=0$ ground state subspace has the same effect, up to an unimportant overall phase, as does applying the projection operator $\Pi^{(12)}_{b_{12}}$ and dividing by the (state-independent) renormalizing factor $\left[ F^{a_{1} a_{2} a_{3} }_{c} \right]^{\ast}_{b_{12} b_{23}}$. In other words, the effect of time evolution (from $t=0$ to $\tau$) under this adiabatic process on a $t=0$ ground state is \emph{exactly} the same as the effect of performing a projective topological charge measurement of the collective charge of anyons $1$ and $2$ with predetermined measurement outcome $b_{12}$. This is shown schematically in Fig.~\ref{fig:proj_via_int}. Operationally, this is identical to the ``forced measurement'' protocol~\cite{Bonderson08a,Bonderson08b}, which allows one to effectively perform a topological charge measurement with predetermined measurement outcome (in certain situations). In hindsight, this should perhaps not be so surprising, since the adiabatic evolution of ground states includes an implicit continual projection into the instantaneous ground state subspace and can be thought of as the continuum limit of a series of measurements, with the final measurement being a projection into the final ground state subspace.

It is always possible to write a Hamiltonian $H(t)$ that satisfies the enumerated properties $1-3$, since one can write a projector $\Pi_{0}(t)$ onto a one-dimensional subspace of $\mathcal{V}^{a_1 a_2 a_3}_{c}$ that interpolates between the initial and final ground state subspaces, such as
\begin{eqnarray}
&& \Pi_{0} (t) = \left( \frac{t}{\tau}  \right)^2 \Pi^{(12)}_{b_{12}} \,\, \Pi^{(123)}_{c} \,\, \Pi^{(12)}_{b_{12}} + \left( \frac{t}{\tau} \right) \left(1 - \frac{t}{\tau} \right) \times \notag \\
&& \left( \left| a_1 , a_2 ; b_{12} \right\rangle \left| b_{12} , a_{3} ; c \right\rangle  \left\langle a_{1}, b_{23} ; c \right| \left\langle a_2 , a_3 ; b_{23} \right| \phantom{\frac{t}{\tau}} \right. \notag \\
&& \left. \phantom{\frac{t}{\tau}} + \left| a_2 , a_3 ; b_{23} \right\rangle \left| a_{1} , b_{23} ; c \right\rangle  \left\langle b_{12}, a_{3} ; c \right| \left\langle a_1 , a_2 ; b_{12} \right| \right) \notag \\
&& \qquad + \left(1- \frac{t}{\tau}  \right)^2 \Pi^{(23)}_{b_{23}} \,\, \Pi^{(123)}_{c} \,\, \Pi^{(23)}_{b_{23}}
.
\end{eqnarray}
However, it is worth considering Hamiltonians that are physically more natural and amenable to experimental implementation. A simple and natural suggestion is to use the linear interpolation
\begin{equation}
H \left( t \right) = \left( \frac{t}{\tau} \right) V^{(12)} + \left(1 - \frac{t}{\tau} \right) V^{(23)}
.
\end{equation}
This Hamiltonian automatically satisfies properties $1$ and $2$. However, it is complicated to determine whether it also satisfies property 3 for general pairwise interactions $V^{(12)}$ and $V^{(23)}$ (unless $\mathcal{V}^{a_1 a_2 a_3}_{c}$ is two-dimensional). In the simple (but non-generic) case where the interactions are given by
\begin{equation}
\label{eq:V_jk_rank_1}
V^{(jk)} = \varepsilon_{jk} \left[ \openone - 2 \Pi^{(jk)}_{b_{jk}} \right]
\end{equation}
with $\varepsilon_{jk}>0$, property $3$ will be satisfied iff $ \left[ F^{a_{1} a_{2} a_{3} }_{c} \right]_{b_{12} b_{23}}  \neq 0$ (which ensures that the projectors $\Pi^{(jk)}_{b_{jk}}$ are not orthogonal). I expect (though have not shown) that property $3$ will also be satisfied for general interactions iff $ \left[ F^{a_{1} a_{2} a_{3} }_{c} \right]_{b_{12} b_{23}}  \neq 0$. For the cases of greatest physical interest, property $3$ is satisfied for arbitrary nontrivial pairwise interactions, because their state spaces $\mathcal{V}^{a_1 a_2 a_3}_{c}$ are two-dimensional [and so reduce to the case in Eq.~(\ref{eq:V_jk_rank_1})] and have $\left[ F^{a_{1} a_{2} a_{3} }_{c} \right]_{b_{12} b_{23}} \neq 0$.

\section{Anyonic Teleportation and Braiding}

Having established that adiabatic manipulation of interactions can be used to produce a forced measurement operation, it is trivial to use it for anyonic teleportation, braiding, and MOTQC in precisely the same way as detailed in Refs.~\onlinecite{Bonderson08a,Bonderson08b}. In particular, one merely needs to consider the case with $a_1 = a_3 = c = a$, $a_2 = \bar{a}$, and $b_{12} = b_{23} = I$. (I note that $\left| \left[ F^{a \bar{a} a }_{a} \right]_{I I} \right| = 1/d_a \neq 0$.)

It is, however, worth reconsidering the use of measurements or forced measurements more generally in these contexts to understand how broadly the methods apply. To this end, I will now examine the case when the topological charge values of the measurement outcomes are not necessarily always the trivial charge $I$. When a particular outcome is necessary or desirable, it is understood that one may use a forced measurement to produce this outcome.

\subsection{Anyonic Teleportation}
\label{sec:Teleportion}

For anyonic teleportation, one considers an anyonic state $\Psi$ partially encoded in anyon $1$ and an ancillary pair of anyons $2$ and $3$, which serve as the entanglement resource. The ancillary anyons are initially in a state with definite fusion channel $b_{23}$ (which must be linked to other anyons, which I denote $\mathcal{A}$, if $b_{23} \neq I$). The combined initial state is written diagrammatically as
\begin{equation}
\left| \Psi \left(a_1,\ldots \right) \right\rangle \left| \mathcal{A} \left(a_2,a_3,\ldots \right) \right\rangle =
\pspicture[shift=-1.0](-1.2,-0.8)(1.2,1.5)
  \psframe[linewidth=0.9pt,linecolor=black,border=0](-1.1,-0.6)(-0.5,0.0)
  \rput[bl]{0}(-0.95,-0.4){$\Psi$}
  \psframe[linewidth=0.9pt,linecolor=black,border=0](0.1,-0.6)(0.7,0.0)
  \rput[bl]{0}(0.25,-0.4){$\mathcal{A}$}
  \psset{linewidth=0.9pt,linecolor=black,arrowscale=1.5,arrowinset=0.15}
  \psline(-0.8,0.0)(-0.8,1.0)
  \psline(0.4,0.0)(0.4,0.5)
  \psline(0.4,0.5)(0.0,1.0)
  \psline(0.4,0.5)(0.8,1.0)
    \psline{->}(-0.8,0.0)(-0.8,0.875)
    \psline{->}(0.4,0.0)(0.4,0.375)
    \psline{->}(0.4,0.5)(0.1,0.875)
    \psline{->}(0.4,0.5)(0.7,0.875)
  \rput[bl]{0}(-0.95,1.1){$a_{1}$}
  \rput[bl]{0}(-0.25,1.1){$a_{2}$}
  \rput[bl]{0}(0.6,1.1){$a_{3}$}
  \rput[bl]{0}(0.55,0.15){$b_{23}$}
 \endpspicture
\end{equation}
where the boxes are used to indicate the encoding details of the states, including other anyons (denoted as ``$\ldots$'') that comprise them.

\begin{figure}[t!]
\begin{center}
  \includegraphics[scale=0.3]{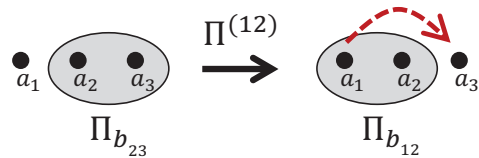}
  \caption{For an anyonic state initially encoded in anyon $1$ and an ancillary pair of anyons $2$ and $3$ in an Abelian fusion channel $b_{23}$, application of a topological charge projector $\Pi^{(12)}_{b_{12}}$, with $b_{12}$ Abelian, teleports the anyonic state information from anyon $1$ to anyon $3$ (indicated by the dashed arrow), while making anyons $1$ and $2$ the new ancillary pair. This projector may be generated via measurements or interactions.}
  \label{fig:teleport}
\end{center}
\end{figure}

To teleport the state information encoded in anyon $1$ to anyon $3$, one applies a projector $\Pi^{(12)}_{b_{12}}$ to the combined state (and renormalizes), at which point anyons $1$ and $2$ become the ancillary pair. It must further be required that $b_{12}$ and $b_{23}$ are Abelian charges, otherwise it will not be possible to dissociate the state information from the ``ancillary'' anyons. In this case, $a_1 = b_{12} \times \bar{a}_{2}$, $a_3 = b_{23} \times \bar{a}_{2}$, $c = b_{12} \times b_{23} \times \bar{a}_{2}$, and $d_{a_{1}}= d_{a_{2}}= d_{a_{3}}=d_{c}$. The post-projected state is
\begin{eqnarray}
&& \left( \left[ F^{a_{1} a_{2} a_{3} }_{c} \right]^{\ast}_{b_{12} b_{23}} \right)^{-1}  \Pi^{(12)}_{b_{12}} \left| \Psi \left(a_1,\ldots \right) \right\rangle \left| \mathcal{A} \left(a_2,a_3,\ldots \right) \right\rangle \notag \\
&=& e^{i \alpha}
\pspicture[shift=-1.7](-1.4,-0.8)(1.2,3.0)
  \psframe[linewidth=0.9pt,linecolor=black,border=0](-1.1,-0.6)(-0.5,0.0)
  \rput[bl]{0}(-0.95,-0.4){$\Psi$}
  \psframe[linewidth=0.9pt,linecolor=black,border=0](0.1,-0.6)(0.7,0.0)
  \rput[bl]{0}(0.25,-0.4){$\mathcal{A}$}
  \psset{linewidth=0.9pt,linecolor=black,arrowscale=1.5,arrowinset=0.15}
  \psline(-0.8,0.0)(-0.8,1.0)
  \psline(0.4,0.0)(0.4,0.5)
  \psline(0.4,0.5)(0.0,1.0)
  \psline(0.4,0.5)(0.8,1.0)
    \psline{->}(-0.8,0.0)(-0.8,0.875)
    \psline{->}(0.4,0.0)(0.4,0.375)
    \psline{->}(0.4,0.5)(0.1,0.875)
    \psline{->}(0.4,0.5)(0.7,0.875)
  \psline(-0.8,1.0)(-0.4,1.5)
  \psline(0.0,1.0)(-0.4,1.5)
  \psline(-0.4,1.5)(-0.4,2.0)
  \psline(-0.8,2.5)(-0.4,2.0)
  \psline(0.0,2.5)(-0.4,2.0)
  \psline(0.8,1.0)(0.8,2.5)
    \psline{->}(-0.4,1.5)(-0.4,1.875)
    \psline{->}(-0.4,2.0)(-0.1,2.375)
    \psline{->}(-0.4,2.0)(-0.7,2.375)
  \rput[bl]{0}(-0.95,2.6){$a_{1}$}
  \rput[bl]{0}(-0.25,2.6){$a_{2}$}
  \rput[bl]{0}(-1.25,0.6){$a_{1}$}
  \rput[bl]{0}(-0.25,0.6){$a_{2}$}
  \rput[bl]{0}(0.6,2.6){$a_{3}$}
  \rput[bl]{0}(0.55,0.15){$b_{23}$}
   \rput[bl]{0}(-0.25,1.6){$b_{12}$}
 \endpspicture
= e^{i \beta}
\pspicture[shift=-1.7](-1.4,-0.8)(1.2,3.5)
  \psframe[linewidth=0.9pt,linecolor=black,border=0](-1.1,-0.6)(-0.5,0.0)
  \rput[bl]{0}(-0.95,-0.4){$\Psi$}
  \psframe[linewidth=0.9pt,linecolor=black,border=0](0.1,-0.6)(0.7,0.0)
  \rput[bl]{0}(0.25,-0.4){$\mathcal{A}$}
  \psset{linewidth=0.9pt,linecolor=black,arrowscale=1.5,arrowinset=0.15}
  \psline(-0.8,0.0)(-0.8,1.0)
  \psline(0.4,0.0)(0.4,0.5)
  \psline(0.4,0.5)(0.0,1.0)
  \psline(0.4,0.5)(0.8,1.0)
    \psline{->}(-0.8,0.0)(-0.8,0.875)
    \psline{->}(0.4,0.0)(0.4,0.375)
    \psline{->}(0.4,0.5)(0.0,1.0)
    \psline{->}(0.8,1.0)(0.8,1.375)
  \psline(-0.8,1.0)(-0.4,1.5)
  \psline(0.0,1.0)(-0.4,1.5)
  \psline(-0.4,1.5)(0.8,2.5)
  \psline(-0.4,2.5)(0.8,1.5)
  \psline(-0.8,3.0)(-0.4,2.5)
  \psline(0.0,3.0)(-0.4,2.5)
  \psline(0.8,1.0)(0.8,1.5)
  \psline(0.8,2.5)(0.8,3.0)
    \psline{->}(-0.4,2.5)(-0.1,2.875)
    \psline{->}(-0.4,2.5)(-0.7,2.875)
    \psline{->}(-0.4,2.5)(-0.7,2.875)
    \psline{->}(0.8,2.5)(0.8,2.875)
  \psline[border=2.0pt](-0.1,1.75)(0.5,2.25)
  \rput[bl]{0}(-0.95,3.1){$a_{1}$}
  \rput[bl]{0}(-0.25,3.1){$a_{2}$}
  \rput[bl]{0}(-1.25,0.6){$a_{1}$}
  \rput[bl]{0}(-0.35,0.7){$f$}
  \rput[bl]{0}(0.6,3.1){$a_{3}$}
  \rput[bl]{0}(0.55,0.15){$b_{23}$}
   \rput[bl]{0}(0.95,1.1){$b_{12}$}
 \endpspicture
\label{eq:teleportation}
\end{eqnarray}
where $e^{i \alpha}$ and $e^{i \beta}$ are unimportant phases (that are straightforward to compute) and $f =  \bar{b}_{12} \times b_{23}$ is an Abelian charge. While it may at first appear that there is still anyonic entanglement between the topological state encoded in anyon $3$ and the ancillary anyons $1$ and $2$,  I emphasize that this is not actually the case. Specifically, the charge line $f$ does not result in any nontrivial anyonic entanglement, because $f$ is Abelian. One must simply keep track of this Abelian charge $f$ as a modification to subsequent readouts, but it does not alter the encoded information. [It is, of course, more clear when $b_{12} = b_{23}$, and hence $f=I$, to see that there is no anyonic entanglement associated with this charge line, since then the final state can be written as $\left| \Psi \left(a_3,\ldots \right) \right\rangle \left| \mathcal{A} \left(a_1,a_2,\ldots \right) \right\rangle $.] The braiding between the $a_3$ and $b_{12}$ charge lines is similarly unimportant (and can also be replaced with a clockwise, rather than counterclockwise braiding), since $b_{12}$ is Abelian, and so the braiding can only contribute an unimportant overall phase. Thus, in this post-projected state, the anyonic state $\Psi$ is partially encoded in anyon $3$ (up to unimportant Abelian factors), while anyons $1$ and $2$ form an ancillary pair that is uncorrelated with $\Psi$, so this is an anyonic teleportation. The planar representation of this is shown in Fig.~\ref{fig:teleport}.

\subsection{Braiding}
\label{sec:Braiding}

\begin{figure}[t!]
\begin{center}
  \includegraphics[scale=0.11]{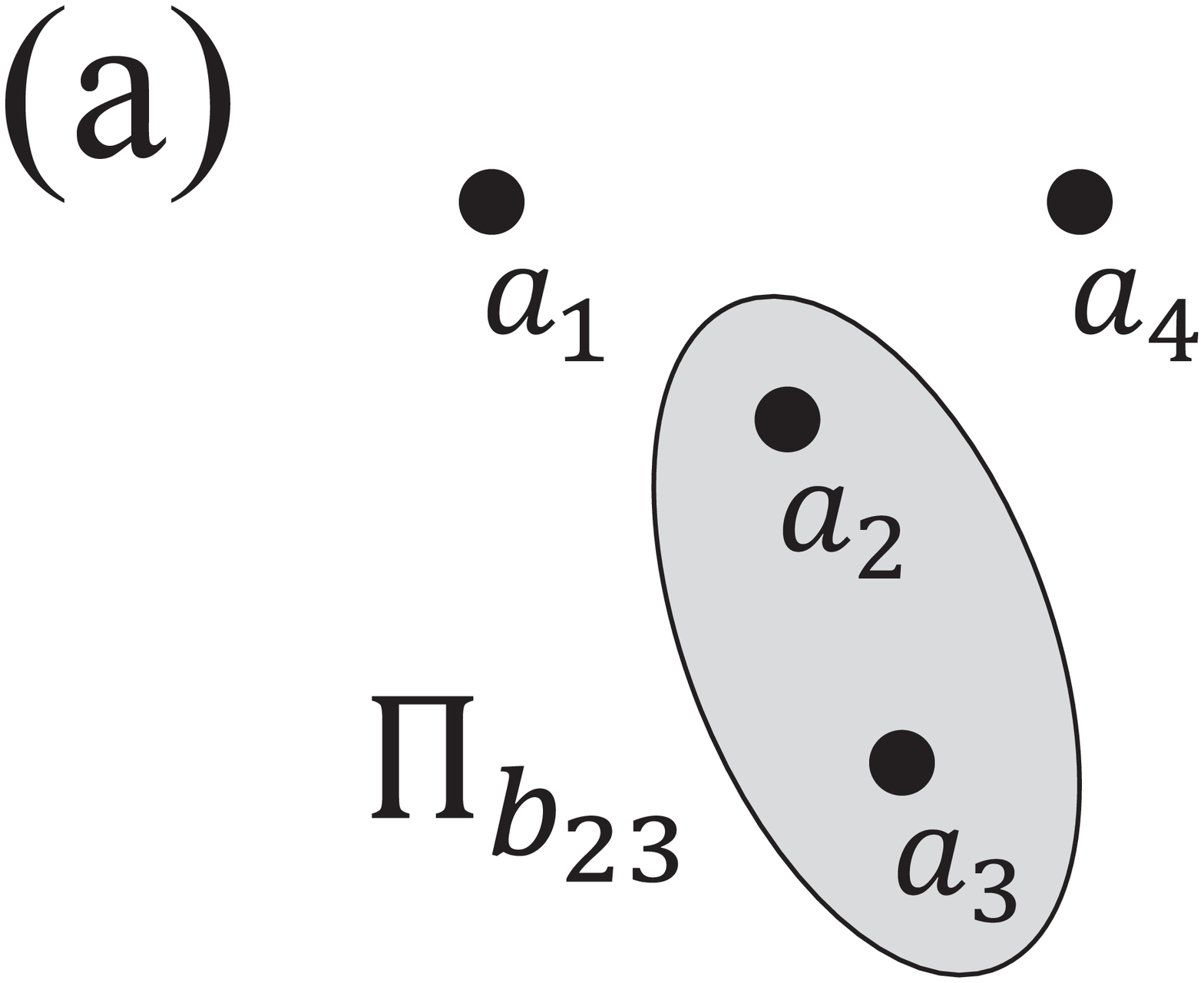}
  \includegraphics[scale=0.11]{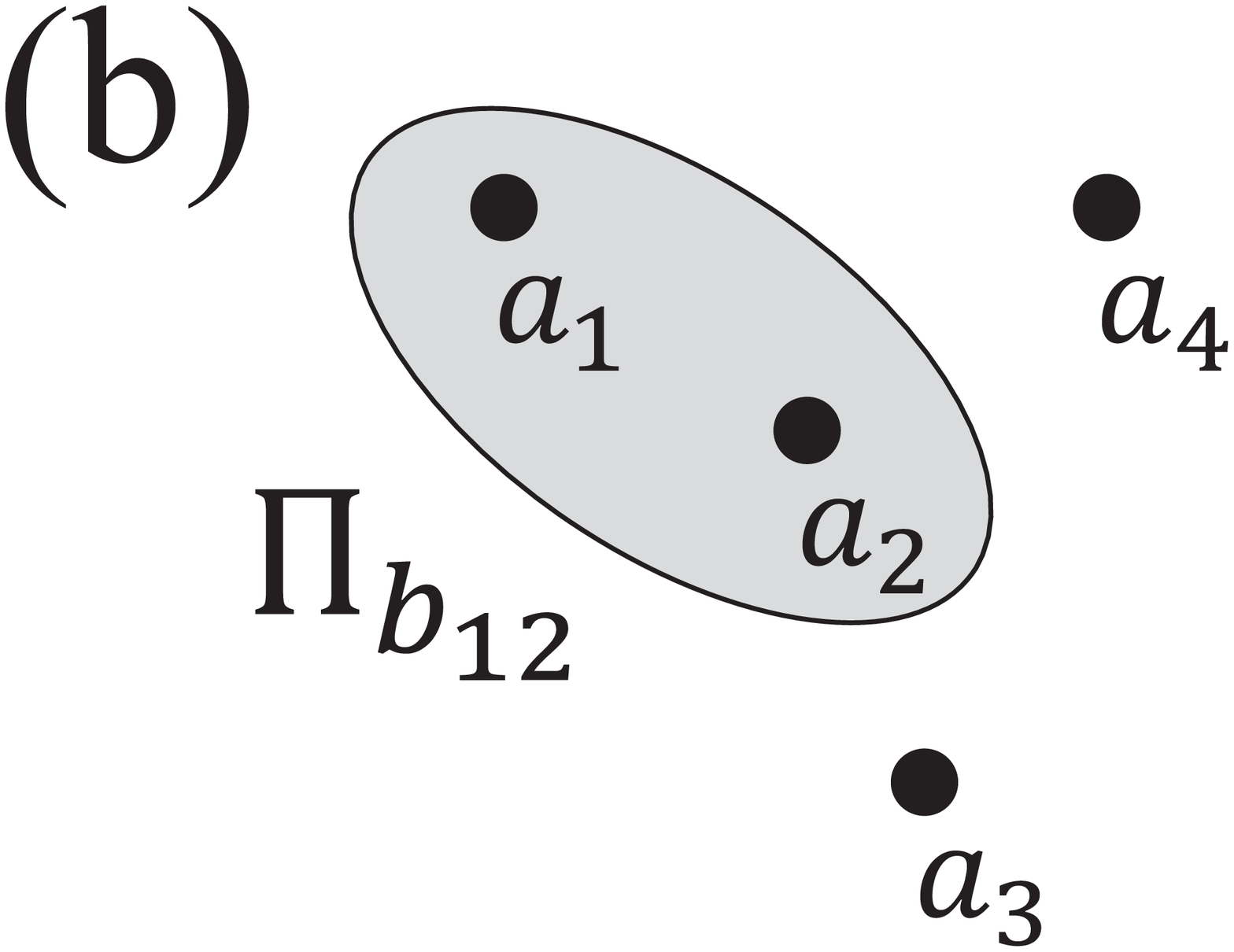}
  \includegraphics[scale=0.11]{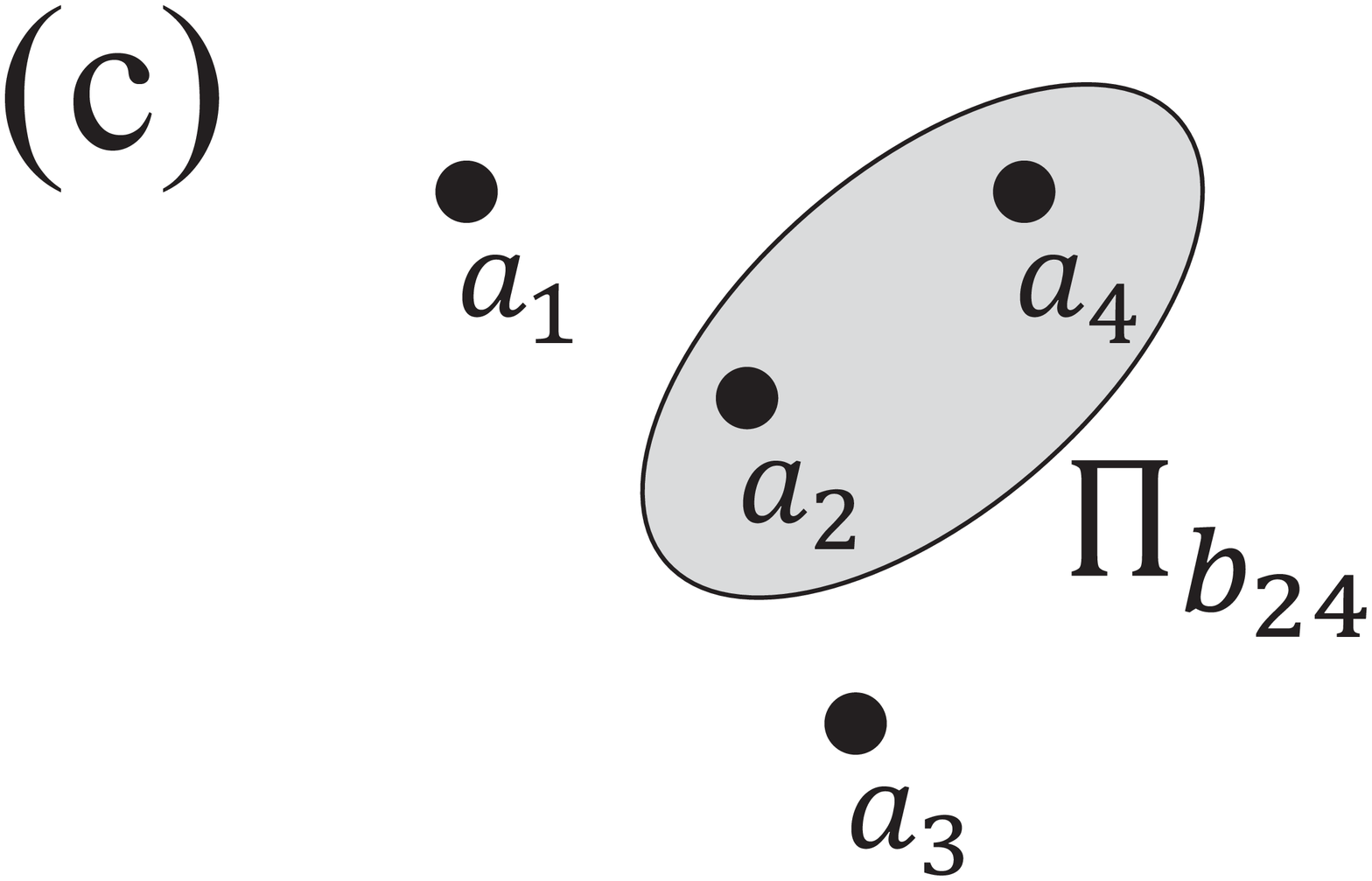}
  \caption{Collective topological charge projectors of pairs of anyons (a) $2$ and $3$, (b) $1$ and $2$, and (c) $2$ and $4$. These projectors may be generated via measurements or interactions and can be used to generate braiding transformations.}
  \label{fig:measurements}
\end{center}
\end{figure}

I now consider four anyons, where anyons $2$ and $3$ are again an ancillary pair and the goal is to implement a braiding transformation for anyons $1$ and $4$, without moving them. I assume anyons $2$ and $3$ are initialized in the fusion channel $b_{23}$ (e.g. by applying a projector). Then I apply a series of pairwise topological charge projections (by performing measurements or forced measurements), first projecting anyons $1$ and $2$ into the fusion channel $b_{12}$, next projecting anyons $2$ and $4$ into the fusion channel $b_{24}$, and finally projecting anyons $2$ and $3$ into the fusion channel $b_{23}^{\prime}$. The configuration of the anyons and pairwise projections is significant for the details of the resulting operator, so for the analysis here I assume the configuration shown in Fig.~\ref{fig:measurements}. The resulting operator is obtained by taking the product of projectors (and dividing by a normalization factor)
\begin{eqnarray}
X &=& C \, \, \Pi^{(23)}_{b^{\prime}_{23}} \Pi^{(24)}_{b_{24}} \Pi^{(12)}_{b_{12}} \Pi^{(23)}_{b_{23}} \notag \\
&=& C^{\prime}
\psscalebox{1}{
\pspicture[shift=-3.7](-1.4,-1.0)(2.0,6.0)
  \psset{linewidth=0.9pt,linecolor=black,arrowscale=1.5,arrowinset=0.15}
  \psline(-0.8,-0.5)(-0.8,1.0)
  \psline(0.4,0.0)(0.4,0.5)
  \psline(0.4,0.5)(0.0,1.0)
  \psline(0.4,0.5)(0.8,1.0)
  \psline(0.4,0.0)(0.8,-0.5)
  \psline(0.4,0.0)(0.0,-0.5)
  \psline(1.6,-0.5)(1.6,1.0)
    \psline{->}(0.4,0.0)(0.4,0.375)
    \psline{->}(0.4,0.5)(0.1,0.875)
    \psline{->}(0.4,0.5)(0.7,0.875)
    \psline{-<}(0.4,0.0)(0.1,-0.375)
    \psline{-<}(0.4,0.0)(0.7,-0.375)
    \psline{-<}(-0.8,0.0)(-0.8,-0.375)
    \psline{-<}(1.6,0.0)(1.6,-0.375)
  \psline(-0.8,1.0)(-0.4,1.5)
  \psline(0.0,1.0)(-0.4,1.5)
  \psline(-0.4,1.5)(-0.4,2.0)
  \psline(-0.8,2.5)(-0.4,2.0)
  \psline(0.0,2.5)(-0.4,2.0)
  \psline(0.8,1.0)(0.8,2.5)
  \psline(1.6,1.0)(1.6,2.5)
    \psline{->}(-0.4,1.5)(-0.4,1.875)
    \psline{->}(-0.4,2.0)(-0.1,2.375)
    \psline{->}(-0.4,2.0)(-0.7,2.375)
  \psline(-0.8,2.5)(-0.8,4.0)
  \psline(0.0,2.5)(1.2,3.0)
  \psline(1.6,2.5)(1.2,3.0)
  \psline(1.2,3.0)(1.2,3.5)
  \psline(0.0,4.0)(1.2,3.5)
  \psline(1.6,4.0)(1.2,3.5)
    \psline{->}(1.2,3.5)(0.3,3.875)
    \psline{->}(1.2,3.5)(1.5,3.875)
    \psline{->}(1.2,3.0)(1.2,3.375)
  \psline[border=2.0pt](0.8,2.5)(0.8,4.0)
  \psline(-0.8,4.0)(-0.8,5.5)
  \psline(1.6,4.0)(1.6,5.5)
  \psline(0.4,4.5)(0.4,5.0)
  \psline(0.8,4.0)(0.4,4.5)
  \psline(0.0,4.0)(0.4,4.5)
  \psline(0.8,5.5)(0.4,5.0)
  \psline(0.0,5.5)(0.4,5.0)
    \psline{->}(0.4,4.5)(0.4,4.875)
    \psline{->}(0.4,5.0)(0.1,5.375)
    \psline{->}(0.4,5.0)(0.7,5.375)
    \psline{->}(-0.8,5.0)(-0.8,5.375)
    \psline{->}(1.6,5.0)(1.6,5.375)
  \rput[bl]{0}(-0.95,-0.8){$a_{1}$}
  \rput[bl]{0}(-0.25,-0.8){$a_{2}$}
  \rput[bl]{0}(0.7,-0.8){$a_{3}$}
  \rput[bl]{0}(1.5,-0.8){$a_{4}$}
  \rput[bl]{0}(0.55,0.05){$b_{23}$}
  \rput[bl]{0}(-0.25,0.6){$a_{2}$}
  \rput[bl]{0}(0.7,0.6){$a_{3}$}
  \rput[bl]{0}(-0.25,1.55){$b_{12}$}
  \rput[bl]{0}(-0.95,2.0){$a_{1}$}
  \rput[bl]{0}(-0.15,2.0){$a_{2}$}
  \rput[bl]{0}(1.35,3.0){$b_{24}$}
  \rput[bl]{0}(1.5,3.55){$a_{4}$}
  \rput[bl]{0}(0.05,3.55){$a_{2}$}
  \rput[bl]{0}(0.55,4.55){$b_{23}^{\prime}$}
  \rput[bl]{0}(-0.95,5.6){$a_{1}$}
  \rput[bl]{0}(-0.25,5.6){$a_{2}$}
  \rput[bl]{0}(0.7,5.6){$a_{3}$}
  \rput[bl]{0}(1.5,5.6){$a_{4}$}
 \endpspicture
}
\end{eqnarray}
where $C$ and $C^{\prime}$ are constants that give the proper normalizations.

It is, again, necessary to require $b_{12}$, $b_{23}$, $b_{23}^{\prime}$, and $b_{24}$ to be Abelian charges. Otherwise, it would not be possible to ensure that the collective topological charge of each $3$-tuple of anyons involved in each teleportation step has definite value ($c_{123}$, $c_{124}$, and $c_{234}$, respectively), which is necessary to apply the results of Sec.~\ref{sec:Interaction_Forced_Measurement}, and to ensure that the resulting operator is unitary. Moreover, if either $b_{23}$ or $b_{23}^{\prime}$ is non-Abelian, it will not be possible to dissociate the operation on anyons $1$ and $4$ from the ``ancillary'' anyons $2$ and $3$.

It is useful (and often natural), though not necessary, to also have $b_{23}=b_{23}^{\prime}$, otherwise there will be an Abelian charge line $f = b_{23} \times \bar{b}_{23}^{\prime}$ connecting the ancillary anyons to the operator, which makes the situation slightly more complicated (though still manageable). Focusing on this case, one finds that $a_1 = b_{12} \times \bar{a}_2$, $a_4 = b_{24} \times \bar{a}_2$, and the $b_{23}= b_{23}^{\prime}$ charge lines can be recoupled and fully dissociated from the operation on anyons $1$ and $4$, so that the operator takes the form
\begin{equation}
X = \hat{X}^{(14)} \otimes \Pi^{(23)}_{b_{23}}
,
\end{equation}
where the operator on anyons $1$ and $4$ is
\begin{eqnarray}
\hat{X}^{(14)} &=& e^{i \phi}
\pspicture[shift=-1.3](-1.4,-0.7)(2.0,2.0)
  \psset{linewidth=0.9pt,linecolor=black,arrowscale=1.5,arrowinset=0.15}
  \psline(-0.8,0.0)(1.6,1.5)
  \psline(1.6,0.0)(-0.8,1.5)
  \psline(-0.2,0.375)(-0.2,1.125)
  \psline(1.0,0.375)(1.0,1.125)
     \psline{->}(0.0,0.5)(0.8,1.0)
     \psline{->}(0.8,0.5)(0.0,1.0)
     \psline{->}(0.8,1.0)(1.4,1.375)
     \psline{->}(0.0,1.0)(-0.6,1.375)
     \psline{-<}(0.8,0.5)(1.4,0.125)
     \psline{-<}(0.0,0.5)(-0.6,0.125)
     \psline{->}(-0.2,0.375)(-0.2,0.875)
     \psline{->}(1.0,0.375)(1.0,0.875)
  \psline[border=2.0pt](0.0,0.5)(0.6,0.875)
  \rput[bl]{0}(-0.95,-0.3){$a_{1}$}
  \rput[bl]{0}(1.5,-0.3){$a_{4}$}
  \rput[bl]{0}(-0.95,1.6){$a_{1}$}
  \rput[bl]{0}(1.5,1.6){$a_{4}$}
  \rput[bl]{0}(-0.05,1.1){$\bar{a}_{2}$}
  \rput[bl]{0}(0.45,1.1){$\bar{a}_{2}$}
  \rput[bl]{0}(-0.8,0.65){$b_{12}$}
  \rput[bl]{0}(1.15,0.65){$b_{24}$}
 \endpspicture
\label{eq:X_1}
\\
&=& e^{i \phi^{\prime}}
\pspicture[shift=-1.3](-1.4,-0.7)(2.0,2.0)
  \psset{linewidth=0.9pt,linecolor=black,arrowscale=1.5,arrowinset=0.15}
  \psline(-0.8,0.0)(1.6,1.5)
  \psline(1.6,0.0)(-0.8,1.5)
  \psline(-0.2,0.375)(-0.2,1.125)
     \psline{->}(0.8,1.0)(1.4,1.375)
     \psline{->}(0.0,1.0)(-0.6,1.375)
     \psline{-<}(0.8,0.5)(1.4,0.125)
     \psline{-<}(0.0,0.5)(-0.6,0.125)
     \psline{->}(-0.2,0.375)(-0.2,0.875)
  \psline[border=2.0pt](0.0,0.5)(0.6,0.875)
  \rput[bl]{0}(-0.95,-0.3){$a_{1}$}
  \rput[bl]{0}(1.5,-0.3){$a_{4}$}
  \rput[bl]{0}(-0.95,1.6){$a_{1}$}
  \rput[bl]{0}(1.5,1.6){$a_{4}$}
  \rput[bl]{0}(-0.5,0.65){$g$}
 \endpspicture
\label{eq:X_2}
\\
\label{eq:X_3}
&=& e^{i \phi^{\prime \prime}} \sum_{c}  \left[ F^{a_{4} g a_{4} }_{c} \right]_{a_{1} a_{1}} R^{a_{4} a_{1}}_{c} \,\, \Pi^{(14)}_{c} \\
\label{eq:X_4}
&=& e^{i \phi } \sum_{c}  R^{\bar{a}_{2} \bar{a}_{2}}_{\hat{c} } \,\, \Pi^{(14)}_{c}
\end{eqnarray}
where $g = b_{12} \times \bar{b}_{24}$, $\hat{c} = c \times \bar{b}_{12} \times \bar{b}_{24}$, and $e^{i \phi}$, $e^{i \phi^{\prime}}$, and $e^{i \phi^{\prime \prime}}$ are unimportant overall phase factors (which may depend on $b_{12}$ and $b_{24}$).

It should be clear that $\hat{X}^{(14)}$ is a modified braiding transformation, with the precise modification depending on $a_j$, $b_{12}$, and $b_{24}$. Furthermore, if $b_{12}=b_{24}$, then $g=0$ and $\hat{X}^{(14)} = e^{i \phi^{\prime \prime}} R_{a_1 a_4}$ is exactly equal to the usual braiding transformation (up to an unimportant overall phase) obtained by exchanging anyons $1$ and $4$ in a counterclockwise fashion. It would be interesting to determine whether these modified braiding operations of Eqs.~(\ref{eq:X_1})--(\ref{eq:X_4}) can augment the computational power of anyons models that do not have computationally universal braiding operations. This is clearly not the case for an anyon model if the permutation of topological charge values given by $\hat{c} = c \times \bar{b}_{12} \times \bar{b}_{24}$ can be obtained from braiding operations. (For Ising anyons, this permutation is a $\sigma_x$ gate and can be obtained by braiding, so these modifications do not augment the computational power, as will be explained in more detail in the next section.)

\section{Ising Anyons and Majorana Fermion Zero Modes}
\label{sec:Ising}

In this section, I consider these results in more detail for Ising anyons, because they are an especially physically relevant example. Ising-type anyons occur as quasiparticles in a number of quantum Hall states~\cite{Moore91,Lee07,Levin07,Bonderson07d,Bonderson09a,Bonderson10b,Bonderson11} that are strong candidates for describing experimentally observed quantum Hall plateaus in the second Landau level~\cite{Willett87,Pan99,Eisenstein02,Xia04,Kumar10}, most notably for the $\nu=\frac{5}{2}$ plateau, which has experimental evidence favoring a non-Abelian state~\cite{Radu08,Willett09a,Willett12}. Ising anyons also describe the Majorana fermion zero modes~\footnote{Since there are always interactions that may lead to energy splitting, it is more accurate to call these ``Majorana $\varepsilon$ modes'' where $\varepsilon$ goes to zero as $\varepsilon = O(e^{-L/\xi})$ for separations $L$ and correlation length $\xi$.}, which exist in vortex cores of two-dimensional (2D) chiral $p$-wave superfluids and superconductors~\cite{Read00,Volovik99}, at the ends of Majorana nanowires (one-dimensional spinless, $p$-wave superconductors)~\cite{Kitaev01,Lutchyn10,Oreg10,Alicea11}, and quasiparticles in various proposed superconductor heterostructures~\cite{Fu08,Sau10a,Alicea10a}. Recently, there have been several experimental efforts to produce Majorana nanowires~\cite{Mourik12,Deng12,Rokhinson12,Das12}.

The Ising anyon model is described by:
\begin{equation*}
\begin{tabular}{|l|}
\hline
$\mathcal{C}=\left\{I,\sigma,\psi \right\}, \quad I\times a=a\times I=a,\quad \psi \times \psi=I,$ \\
$\qquad \qquad \sigma \times \psi=\psi \times \sigma=\sigma,\quad \sigma \times \sigma=I+\psi$ \\ \hline
\qquad \qquad$\left[ F_{\sigma}^{\sigma \sigma \sigma}\right] _{ef}=
\left[
\begin{array}{rr}
\frac{1}{\sqrt{2}} & \frac{1}{\sqrt{2}} \\
\frac{1}{\sqrt{2}} & \frac{-1}{\sqrt{2}}%
\end{array}\right] _{ef}^{\phantom{T}}$ \\
\qquad \qquad $\left[ F_{\psi}^{\sigma \psi \sigma}\right] _{\sigma \sigma}=%
\left[ F_{\sigma}^{\psi \sigma \psi}\right] _{\sigma \sigma_{\phantom{j}}}\!\!=
-1 $ \\ \hline
\qquad $R_{I}^{\sigma \sigma}=e^{-i\frac{\pi }{8}},\quad R_{\psi}^{\sigma \sigma}=e^{i\frac{3\pi }{8}},$ \\
\qquad $R_{\sigma}^{\sigma \psi}=R_{\sigma}^{\psi \sigma}=-i,\quad R_{I}^{\psi \psi}=-1$ \\ \hline
\qquad $d_{I}=d_{\psi}=1,\quad d_{\sigma_{\phantom{j}}}\!\!=\sqrt{2}$ \\ \hline
\end{tabular}%
\end{equation*}%
where $e,f\in \left\{ I,\psi\right\} $, and only the non-trivial $F$-symbols and $R$-symbols are listed. ($F$-symbols and $R$-symbols not listed are equal to $1$ if their vertices are permitted by the fusion algebra, and equal to $0$ if they are not permitted.) The topological charge $\psi$ corresponds to a fermion, while $\sigma$ corresponds to a non-Abelian anyon. In Majorana fermion systems, the zero modes correspond to the $\sigma$ anyons. In this way, the fusion rule $\sigma \times \sigma = I+\psi$ indicates that a pair of zero modes combines to a fermion mode, which can either be unoccupied or occupied, corresponding to the $I$ or $\psi$ fusion channel, respectively. The braiding operator for exchanging Majorana zero modes is given by the braiding of the $\sigma$ Ising anyons, up to an overall phase ambiguity.

The braiding transformations of Ising anyons are not, by themselves, computationally universal, as they only generate a subset of the Clifford gates. However, they nonetheless provide a topologically protected gate set that is very useful for quantum information processing and error correction~\cite{Bravyi05}.

For anyonic teleportation, one considers the case where $a_1 = a_2 = a_3 = \sigma$. Then, $b_{12}$, $b_{23}$, and $f=\bar{b}_{12} \times b_{23}$ in Eq.~(\ref{eq:teleportation}) can equal either $I$ or $\psi$. When $f=I$, there is no charge line connecting the final ancillary pair of anyons $1$ and $2$, to anyon $3$, so the state information that was initially encoded in anyon $1$ is teleported to anyon $3$, with no modifying factors. When $f=\psi$, the state information is similarly teleported from anyon $1$ to $3$, but the overall anyonic charge of the encoded state $\Psi$ now has an extra fermionic parity $\psi$ associated with anyon $3$, entering through the charge line $f$. The encoded state information is not altered, but if one is attempting to access the state information through a collective topological charge measurement including anyon $3$, then one must remember to factor out this extra fermionic parity when identifying the state's measurement outcome.

For the (modified) braiding transformation generated from measurements or forced measurements, one considers the case when $a_1 = a_2 = a_3 = a_4= \sigma$. Then, $b_{12}$, $b_{23}$, $b_{24}$, and $g={b}_{12} \times \bar{b}_{24}$ in Eqs.~(\ref{eq:X_1})--(\ref{eq:X_4}) can equal either $I$ or $\psi$, and $\tilde{c} = g \times c$. When $g=I$, the operator
\begin{equation}
\hat{X}^{(14)} = e^{i \varphi} R_{\sigma \sigma}
\end{equation}
is equal to the braiding exchange of the two $\sigma$ anyons in a counterclockwise fashion (apart from an unimportant overall phase $e^{i \varphi}$). When $g=\psi$, the operator becomes
\begin{equation}
\hat{X}^{(14)} = e^{i \varphi^{\prime}} R^{-1}_{\sigma \sigma}
,
\end{equation}
which is equal to the braiding exchange of the two $\sigma$ anyons in a \emph{clockwise} fashion (apart from a different unimportant overall phase $e^{i \varphi^{\prime}}$). The modification due to $g=\psi$ effectively reverses the chirality of the braiding exchange.

\section{Majorana Wires}

It is useful and interesting to consider the results of this paper in the context of Majorana nanowires. In particular, in the discretized model of Majorana nanowires, the translocation and exchange of the Majorana zero modes localized at the ends of wires can be understood as applications of anyonic teleportation and measurement-generated braiding transformation, as I now explain.

Kitaev's $N$-site fermionic chain model, for a spinless, $p$-wave superconducting wire is given by the Hamiltonian~\cite{Kitaev01}
\begin{eqnarray}
\label{eq:Majorana_wire}
H &=& -\mu \sum_{j=1}^{N} \left( c_{j}^{\dagger} c_j - \frac{1}{2} \right) - w \sum_{j=1}^{N-1} \left( c_{j}^{\dagger} c_{j+1} + c_{j+1}^{\dagger} c_{j} \right) \notag \\
&& - \sum_{j=1}^{N-1} \left( \Delta c_{j} c_{j+1} + \Delta^{\ast} c_{j+1}^{\dagger} c_{j}^{\dagger} \right)
,
\end{eqnarray}
where $\mu$ is the chemical potential, $w$ is the hopping amplitude, $\Delta = |\Delta | e^{i \theta}$ is the induced superconducting gap, and the $j$th site has (spinless) fermionic annihilation and creation operators, $c_j$ and $c_j^{\dagger}$, respectively. This Hamiltonian exhibits two gapped phases (assuming the chain is long, i.e., $N \gg1$):

(a) The trivial phase with a unique ground state occurs for $2|w|<\mu$.

(b) The non-trivial phase with twofold-degenerate ground states and zero modes localized at the endpoints occurs for $2|w|>\mu$ and $\Delta \neq 0$.

A powerful way of understanding this model comes from rewriting the fermionic operator $c_j$ of each site in terms of two Majorana operators~\cite{Kitaev01}
\begin{eqnarray}
\gamma_{2j-1} &=& e^{i \frac{\theta}{2} } c_j + e^{-i \frac{\theta}{2} } c_{j}^{\dagger} \\
\gamma_{2j} &=& -i e^{i \frac{\theta}{2} } c_j + i e^{-i \frac{\theta}{2} } c_{j}^{\dagger}
.
\end{eqnarray}
In this way, the two gapped phases can be qualitatively understood by considering the following special cases inside each phase:

(a) $\mu<0$ and $w = \Delta =0$, for which the Hamiltonian becomes
\begin{equation}
\label{eq:Majorana_wire_a}
H_a = \left( \frac{-\mu}{2} \right) \sum_{j=1}^{N} i \gamma_{2j-1} \gamma_{2j}
.
\end{equation}

(b) $\mu=0$ and $w = |\Delta| > 0$, for which the Hamiltonian becomes
\begin{equation}
\label{eq:Majorana_wire_b}
H_b = w \sum_{j=1}^{N-1} i \gamma_{2j} \gamma_{2j+1}
.
\end{equation}

\begin{figure}[t!]
\begin{center}
  \includegraphics[scale=0.35]{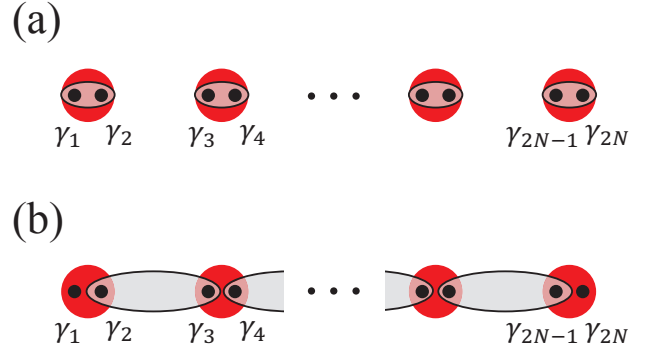}
  \caption{Characteristic pairing of the two gapped phases of Kitaev's fermionic chain model of Eq.(\ref{eq:Majorana_wire}). Fermionic sites (red dots) can be expressed in terms of two Majorana operators $\gamma_{2j-1}$ and $\gamma_{2j}$ (black dots). In the trivial phase (a), the dominant interaction (shaded ovals) is between pairs of Majorana operators $\gamma$ on the same site. In the non-trivial phase (b), the dominant interaction (shaded ovals) is between pairs of Majorana operators on adjacent sites and there is an unpaired Majorana operator localized at each end of the chain.}
  \label{fig:Majorana_chain}
\end{center}
\end{figure}

I note that any pair of Majorana operators $\gamma_j$ and $\gamma_k$ can be written as a fermionic operator $\tilde{c} =\frac{1}{2}\left(\gamma_{j} + i \gamma_k \right)$, in which case $i \gamma_j \gamma_k = 2 \tilde{c}^{\dagger} \tilde{c} - 1$. Thus, the eigenvalue $-1$ of $i \gamma_j \gamma_k $ corresponds to an unoccupied fermionic state, while the $+1$ eigenvalue corresponds to an occupied fermionic state. In $H_a$, each Majorana operator is paired with the other Majorana operator on the same site, such that the fermionic state at each site is unoccupied in the ground state. In $H_b$, each Majorana operator is paired with a Majorana operator in an adjacent site (such that their corresponding fermionic state is unoccupied in the ground states), except for $\gamma_1$ and $\gamma_{2N}$, which are unpaired (i.e., they do not occur in the expression for $H_b$). These unpaired Majorana operators result in zero modes, which give rise to a twofold degeneracy of ground states corresponding to $i \gamma_{1} \gamma_{2N} = \pm 1$.

The pairings exhibited for these two special cases are characteristic of their corresponding phases, as shown in Fig.~\ref{fig:Majorana_chain}. In the phase (a), the dominant interaction is between pairs of Majorana operators on the same site. In the phase (b), the dominant interaction is between pairs of Majorana operators on adjacent sites and there are Majorana zero modes localized at both ends of the chain, giving rise to twofold-degenerate ground states. For the general case in the (b) phase, the ground state degeneracy and zero mode localization is topological, meaning they will generally not be exact, but rather involve corrections that are exponentially suppressed in the length of the chain as $O(e^{-\alpha N})$, for some constant $\alpha$, and they will be robust to deformations of the Hamiltonian that do not close the gap. In other words, they are actually Majorana $\varepsilon$ modes with $\varepsilon=O(e^{-\alpha N})$.

\begin{figure}[t!]
\begin{center}
  \includegraphics[scale=0.3]{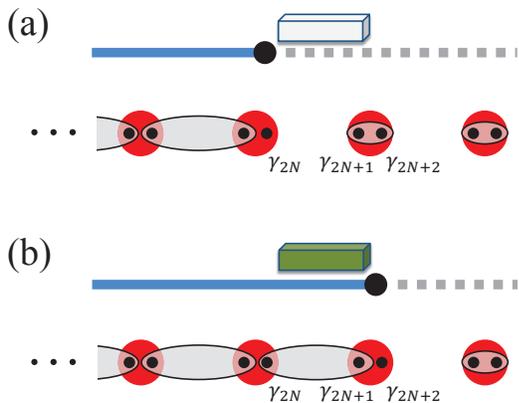}
  \caption{Moving the endpoint of a Majorana wire (solid blue line) and its associated Majorana zero mode (black dot) into a region of topologically trivial wire (dashed grey line) by locally tuning the system, so that a segment of wire changes from the trivial phase (a) to the non-trivial phase (b). The rectangular box represents gates that may be used to locally tune the system. The corresponding configuration of the discretized model is shown in each case.}
  \label{fig:Majorana_chain_extension}
\end{center}
\end{figure}

One can now consider operations that move one of the endpoints of the wires and, hence, the Majorana zero mode localized there, as shown in Fig.~\ref{fig:Majorana_chain_extension}. This can be done by locally tuning the system parameters to extend the topological wire segment into a region of non-topological wire or retract it from a non-topological region. In the discretized model, this amounts to adiabatically tuning the Hamiltonian at the interface of a trivial segment and a nontrivial one, so that a site initially in the (a) phase becomes the new endpoint of the wire in the (b) phase, or vice-versa.

\begin{figure}[t!]
\begin{center}
  \includegraphics[scale=0.16]{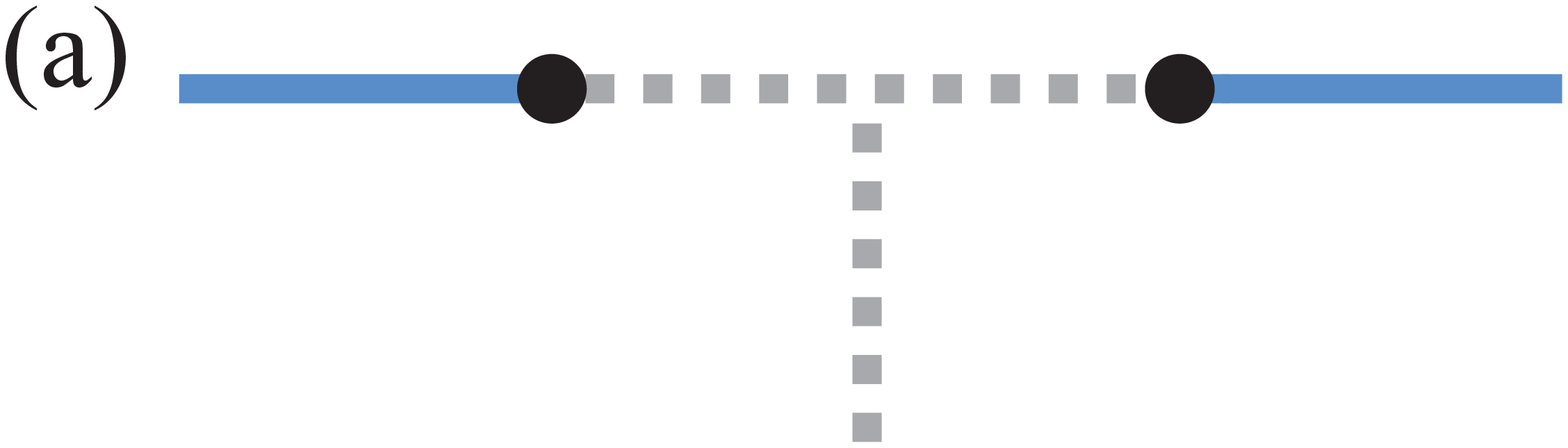}
  \includegraphics[scale=0.16]{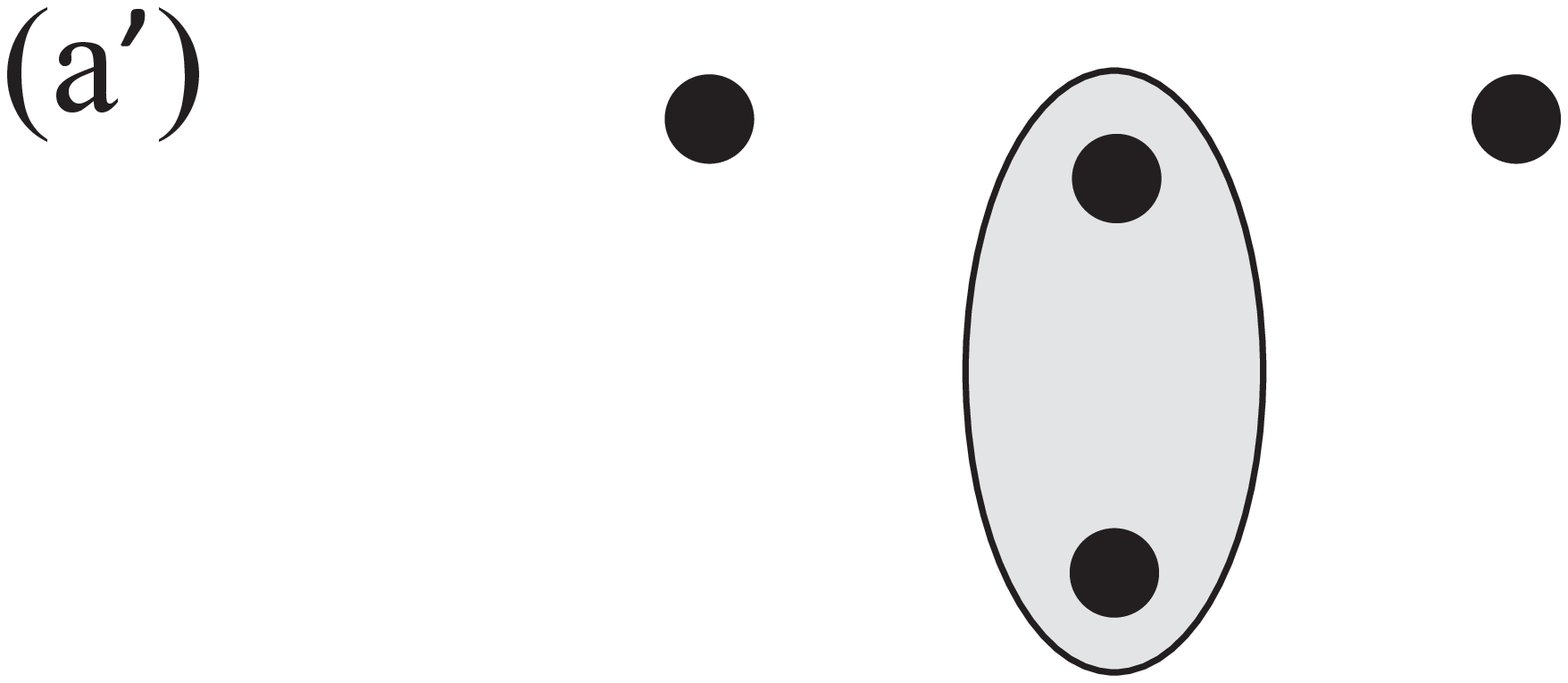}
  \includegraphics[scale=0.16]{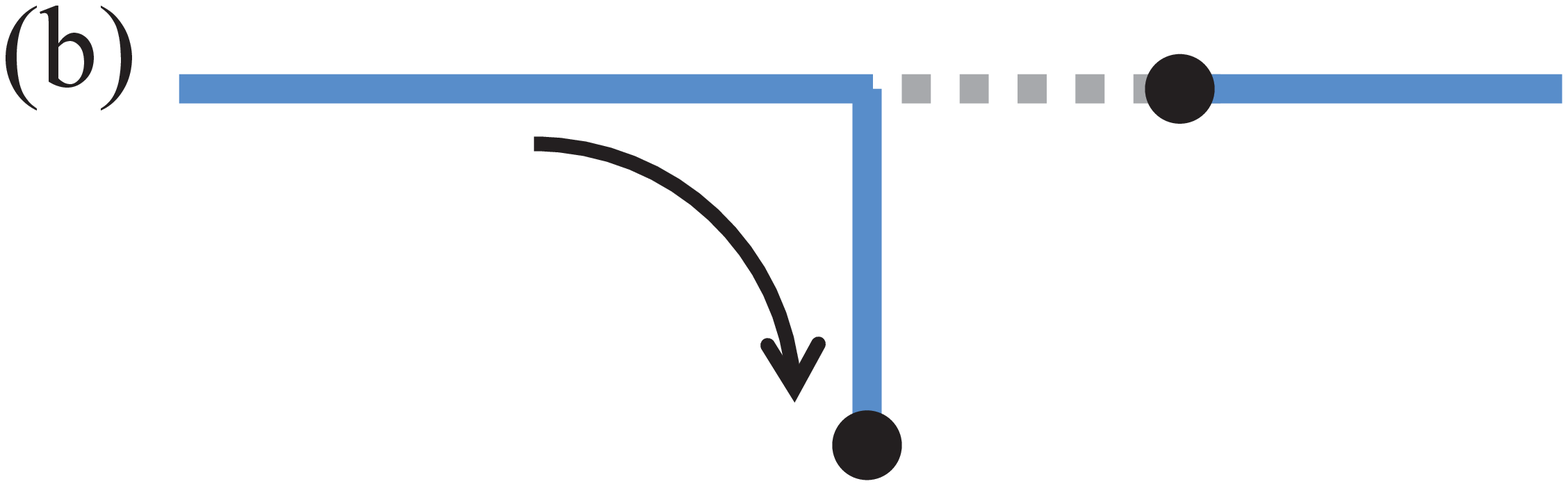}
  \includegraphics[scale=0.16]{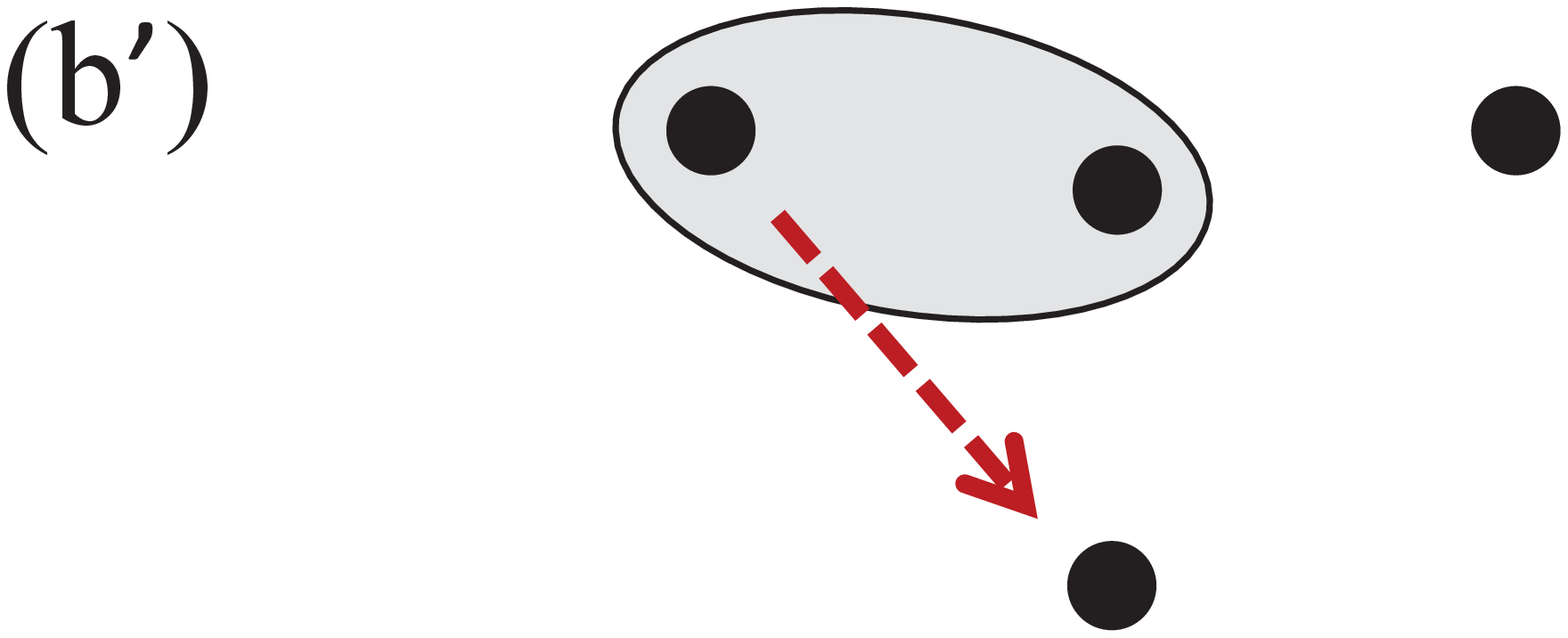}
  \includegraphics[scale=0.16]{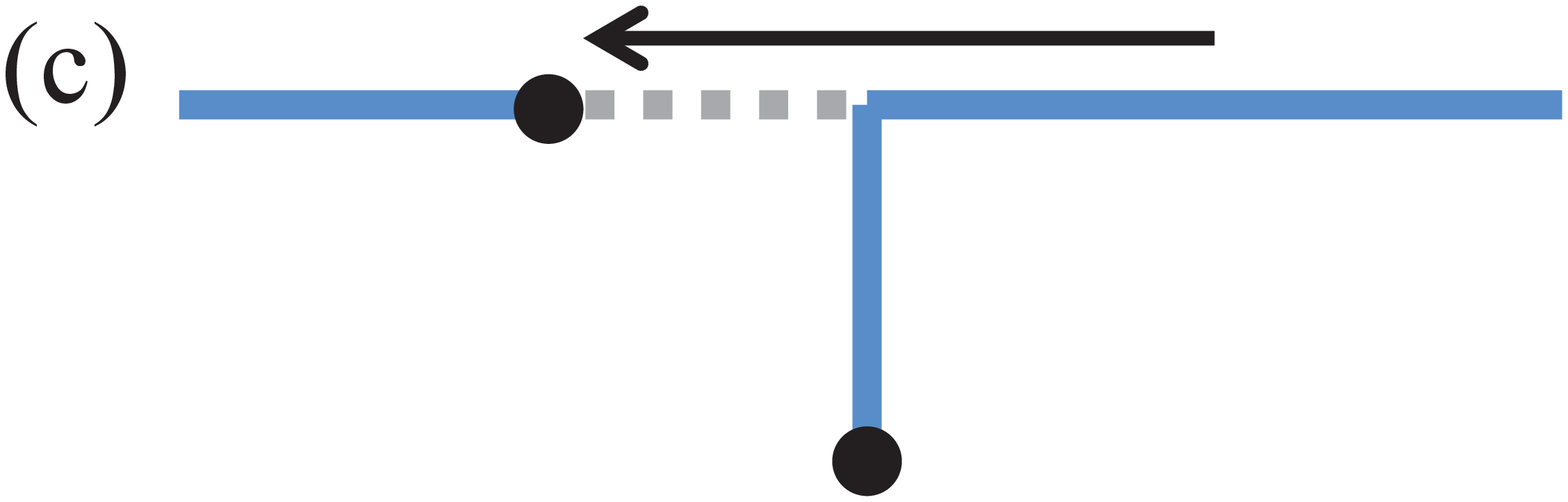}
  \includegraphics[scale=0.16]{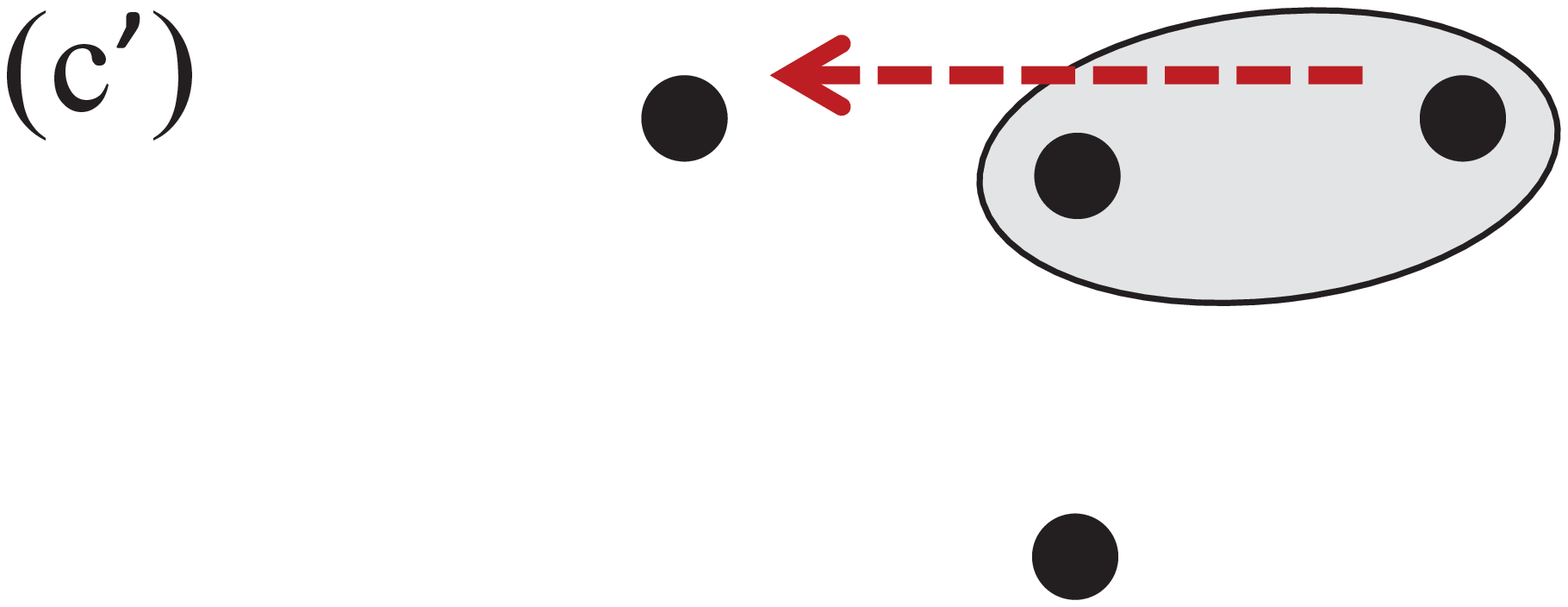}
  \includegraphics[scale=0.16]{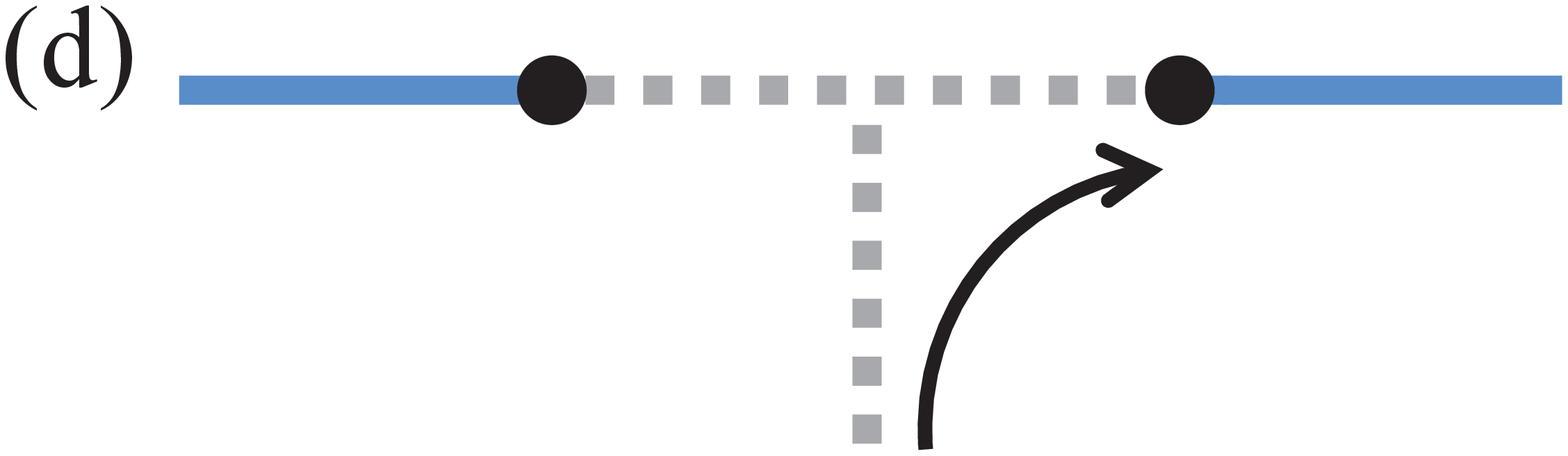}
  \includegraphics[scale=0.16]{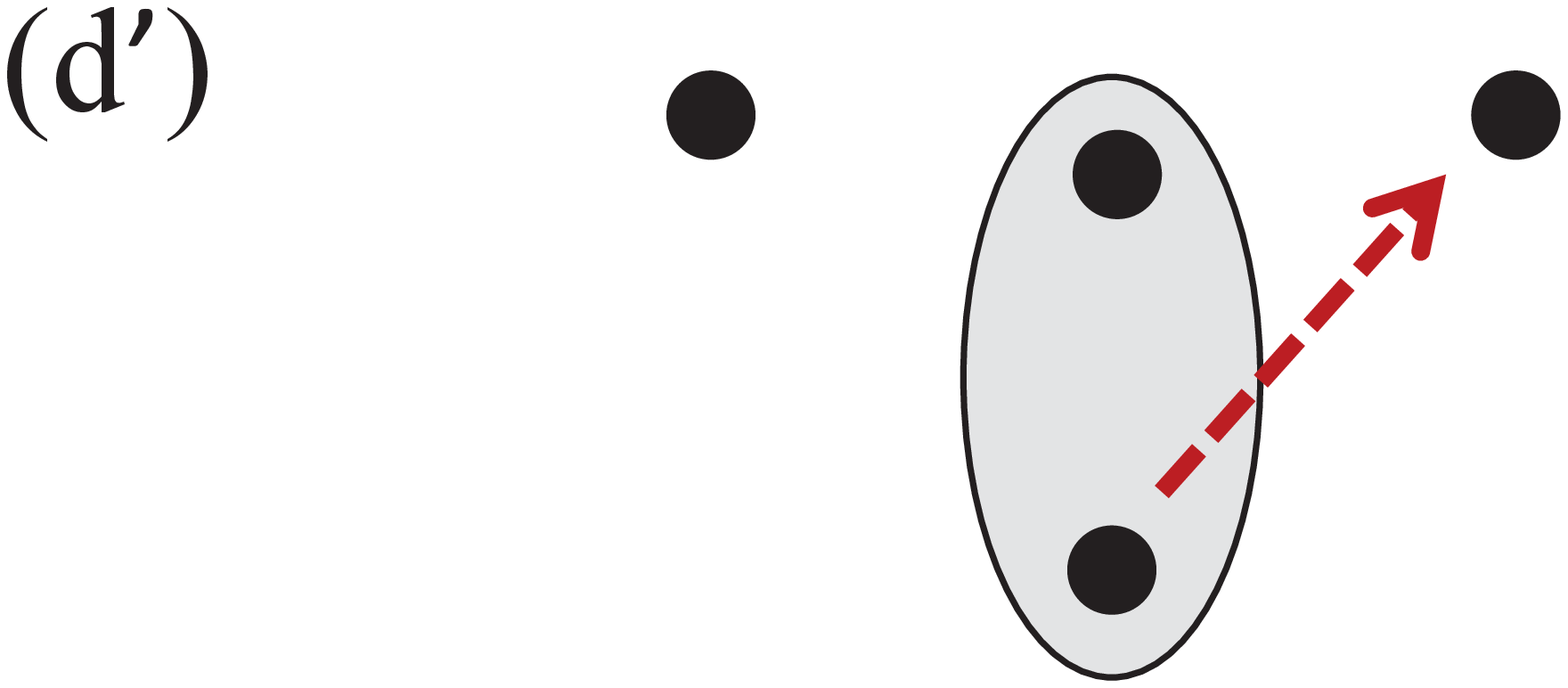}
  \caption{Exchanging the endpoints of Majorana wires (solid blue lines) and their associated Majorana zero modes (black dots) using a T-junction through a series of operations (a)-(d). (Dashed grey lines indicate wire segments in the trivial phase.) Each step in the series can be mapped to a corresponding application of a projector (a')-(d'). The dashed red arrows indicate anyonic teleportation of encoded state information, as in Fig.~\ref{fig:teleport}. This series of projections is exactly the same as that used to generate braiding transformations in Sec.~\ref{sec:Braiding}.}
  \label{fig:wires_braid}
\end{center}
\end{figure}

To be concrete, I consider the Hamiltonian $H(t)$, which acts as $H_b$ on sites $1,\ldots,N$ and as $H_a$ on sites $N+2,N+3,\ldots$, for all $t$, while its time-dependent action on the Majorana operators $\gamma_{2N}$, $\gamma_{2N+1}$, and $\gamma_{2N+2}$ (associated with sites $N$ and $N+1$) is given by
\begin{eqnarray}
H(t) &=& \left( 1 - \frac{t}{\tau} \right) \left[ \left(\frac{-\mu}{2} \right) i \gamma_{2N+1} \gamma_{2N+2} \right] \notag \\
&& + \left( \frac{t}{\tau}\right) \left[ w \,\, i \gamma_{2N} \gamma_{2N+1} \right]
\end{eqnarray}
for $0 \leq t \leq \tau$. This locally takes the Hamiltonian from the form $H_a$ at $t=0$ to $H_b$ at $t=\tau$ on site $N+1$, extending the length of the (b) phase region and moving the zero mode from site $N$ (associated with $\gamma_{2N}$) to site $N+1$ (associated with $\gamma_{2N+2}$). It should be clear that this has exactly the form of time-dependent Hamiltonians satisfying properties $1-3$ described in Sec.~\ref{sec:Interaction_Forced_Measurement}. In particular, using the mapping between Ising anyons and Majorana fermion zero modes explained in Sec.~\ref{sec:Ising}, one can replace the Majorana operators $\gamma_j$ with $\sigma$ Ising anyons, roughly speaking. The unoccupied fermionic state of a pair of Majorana operators corresponds to a pair of $\sigma$ anyons fusing into the $I$ channel and the occupied fermionic state corresponds to them fusing into the $\psi$ channel. The pairwise interaction $i \gamma_j \gamma_{k}$ maps to the interaction $V^{(jk)} = \openone - 2 \Pi^{(jk)}_{I}$ of Ising anyons, which energetically favors the $b_{jk}=I$ fusion channel. Thus, one can view this operation, which extends the Majorana wire and moves the zero mode from site $N$ to site $N+1$, as an anyonic teleportation of the anyonic state information encoded in anyon $2N$ to anyon $2N+2$, as explained in Sec.~\ref{sec:Teleportion}. The ``ancillary anyons'' in this case are drawn from and absorbed into the bulk of the wires. The relation to anyonic teleportation can be seen clearly by comparing the discretized model in Fig.~\ref{fig:Majorana_chain_extension} to Fig.~\ref{fig:teleport}. In order to retract the nontrivial wire segment, one simply needs to run this process in reverse. The (special case) Hamiltonian described in this paragraph provides the cleanest example for changing a segment between the (a) and (b) phases and its relation to anyonic teleportation, but the general case is qualitatively the same.

In Ref.~\onlinecite{Alicea11}, it was shown that with a wire network involving ``T-junctions,'' one could perform a series of operations that exchange the endpoints of Majorana nanowires and, hence, the zero modes localized at them, and that these exchanges would result in transformations equivalent to the braiding transformation of Ising anyons (up to overall phase). It should now be clear that such exchange operations can similarly be viewed as a series of anyonic teleportations that gives rise to (modified) braiding transformations as explained in Sec.~\ref{sec:Braiding}. This relation is shown schematically in Fig.~\ref{fig:wires_braid}. This, in part, explains the observation~\cite{Alicea11,Clarke11a} that an exchange of the endpoints of Majorana nanowires using a T-junction can realize either chirality of Ising braiding transformation, depending on the details of the T-junction, not just on the chirality of the Majorana wires and order of operations. In particular, as shown in Sec.~\ref{sec:Ising}, the chirality of the Ising transformation implemented will depend, in part, on the (forced) measurement outcomes $b_{12}$ and $b_{24}$, which translate into the signs of coupling interactions at the T-junction.

It is straightforward to extend the results of this section (and paper) to the $\mathbb{Z}_n$-Parafendleyon wires~\cite{Clarke12,Lindner12,Cheng12,Fendley12,Vaezi12}. These can be thought of as generalizations of Majorana wires for which the zero modes localized at the endpoints possess $2n$ Abelian fusion channels, rather than two. The results of Eqs.~(\ref{eq:X_1})-(\ref{eq:X_4}) similarly explain the possibility of realizing different transformations when exchanging the zero modes (though in the general case, it is not simply the difference between counterclockwise and clockwise braiding chiralities)~\cite{Clarke12,Lindner12}.

\phantom{Hi}

\acknowledgements
I thank M. Freedman and C. Nayak for useful discussions.
I acknowledge the hospitality and support of the Aspen Center for Physics under the NSF grant No. 1066293.


\end{document}